\documentclass[twocolumn,trackchanges]{aastex62}
\usepackage{placeins}
\usepackage{mathtools, nccmath}
\usepackage{lipsum}
\usepackage{siunitx,etoolbox}
\usepackage{capt-of}

\usepackage{hyperref}
\usepackage{longtable}
\usepackage{float}
\usepackage{placeins}
\usepackage{lineno}

\graphicspath{{./}{figures/}}

\shortauthors{Dumont et al.}

\revised{\today}
\accepted{ January 19, 2022}

\begin{document}

\title{A population of luminous globular clusters and stripped nuclei with elevated mass to light ratios around NGC~5128\footnote{This paper includes data gathered with the $6.5$~m Magellan Telescope at Las Campanas Observatory, Chile}}

\correspondingauthor{Antoine Dumont}
\email{antoine.dumont.neira@hotmail.com}

\author[0000-0003-0234-3376]{Antoine Dumont \href{mailto:antoine.dumont.neira@hotmail.com}}
\affil{Department of Physics and Astronomy, University of Utah\\
115 South 1400 East, Salt Lake City, UT 84112, USA}

\author[0000-0003-0248-5470]{Anil C. Seth \href{mailto:seth@astro.utah.edu}}
\affiliation{Department of Physics and Astronomy, University of Utah\\
115 South 1400 East, Salt Lake City, UT 84112, USA}

\author[0000-0002-1468-9668]{Jay Strader  \href{mailto:strader@pa.msu.edu}}
\affiliation{Department of Physics and Astronomy Michigan State University Biomedical \& Physical Sciences\\567 Wilson Rd, Room 3275 East Lansing, MI 48824-2320}

\author[0000-0001-6215-0950]{Karina Voggel}
\affiliation{Universite de Strasbourg, CNRS, Observatoire astronomique de Strasbourg, UMR 7550, F-67000 Strasbourg, France}

\author[0000-0003-4102-380X]{David J. Sand}
\affil{Steward Observatory, University of Arizona, 933 North Cherry Avenue, Tucson, AZ 85721, USA}

\author[0000-0002-1718-0402]{Allison K. Hughes}
\affil{Steward Observatory, University of Arizona, 933 North Cherry Avenue, Tucson, AZ 85721, USA}

\author[0000-0003-2352-3202]{Nelson Caldwell}
\affiliation{Center for Astrophysics, Harvard \& Smithsonian, 60 Garden Street, Cambridge, MA 02138, USA}

\author[0000-0002-1763-4128]{Denija Crnojevi\'{c}}
\affiliation{University of Tampa, 401 West Kennedy Boulevard, Tampa, FL 33606, USA}


\author{Mario Mateo}
\affiliation{Department of Physics and Astronomy Michigan State University Biomedical \& Physical Sciences\\567 Wilson Rd, Room 3275 East Lansing, MI 48824-2320}

\author[0000-0002-4272-263X]{John I. Bailey, III}
\affiliation{Department of Physics Broida Hall, University of California, Santa Barbara, CA 93106, USA}

\author{Duncan A. Forbes}
\affiliation{Centre for Astrophysics \& Supercomputing, Swinburne University, Hawthorn, VIC 3122, Australia}


\begin{abstract}
The dense central regions of tidally disrupted galaxies can survive as ultra-compact dwarfs (UCDs) that hide among the luminous globular clusters (GCs) in the halo of massive galaxies. An exciting confirmation of this model is the detection of overmassive black holes in the centers of some UCDs, which also lead to elevated dynamical mass-to-light ratios ($M/L_{dyn}$). Here we present new high-resolution spectroscopic observations of 321 luminous GC candidates in the massive galaxy NGC~5128/Centaurus~A. Using these data we confirm 27 new luminous GCs, and measure velocity dispersions for 57 luminous GCs (with $g$-band luminosities between $2.5 \times 10^5$ and $2.5 \times 10^7 L_{\odot}$), of which 48 are new measurements. Combining these data with size measurements from Gaia, we determine the $M/L_{dyn}$ for all 57 luminous GCs. We see a clear bimodality in the $M/L_{dyn}$ distribution, with a population of normal GCs with mean $M/L_{dyn}=1.51\pm0.31$, and a second population of $\sim$20 GCs with elevated mean $M/L_{dyn}=2.68\pm0.22$. We show that black holes with masses $\sim4$--18\% of the luminous GCs can explain the elevated mass-to-light ratios. Hence, it is plausible that the NGC~5128 sources with elevated $M/L_{dyn}$ are mostly stripped galaxy nuclei that contain massive central black holes, though future high spatial resolution observations are necessary to confirm this hypothesis for individual sources. We also present a detailed discussion of an extreme outlier, \textit{VHH81-01}, one of the largest and most massive GC in NGC~5128, making it an exceptionally strong candidate to be a tidally stripped nucleus.

\end{abstract}


\section{Introduction\label{sec:intro}} 

A majority of lower mass galaxies host dense nuclear star clusters \citep{Neumayer2020}. During merging, the tidal forces of a larger galaxy can strip away the contents of a smaller one leaving behind these nuclear star clusters to orbit in the halo of the larger galaxy as a stripped galaxy nucleus \citep[e.g.][]{Pfeffer2013}.
Such stripped galaxy nuclei can be challenging to distinguish from massive globular clusters (GCs).  The collection of massive GCs and stripped galaxy nuclei are often referred to as ultra-compact dwarf galaxies (UCDs). The physical definition of UCDs varies widely in the literature \citep[e.g][]{Mieske2013,Liu2015}, but commonly refers to objects with $\gtrsim 2\times10^{6}$~M$_{\odot}$ and half-light radii of $\gtrsim 7-10$~pc \citep[e.g][]{Mieske2008,Forbes11}.  However, nuclear star clusters extend down to lower masses and sizes than commonly used cutoffs for defining UCDs and can overlap in mass and size with GCs \citep{Neumayer2020}.  Thus the definitions of UCDs neither exclude all massive GCs nor do they include all stripped nuclei.  
Due to the ambiguity in the classification of UCDs, throughout this paper we use the term ``luminous GCs''.  These luminous GCs refer to both massive GCs and stripped galaxy nuclei without assuming a specific mass, luminosity, or size cut.

A large number of stripped nuclei are expected around massive galaxies and in galaxy clusters due to the hierarchical nature of galaxy merging.  Simulations are consistent with most of the higher mass GCs ($\gtrsim$10$^7$~M$_\odot$) in clusters being stripped galaxy nuclei \citep{Pfeffer2016,Mayes2021,Norris2019}. They also suggest that lower mass galaxies will contain lower mass stripped nuclei, and that the Milky Way may contain 2-6 stripped nuclei \citep{Pfeffer2014,Kruijssen2019}.  
%


Stripped nuclei can be distinguished observationally from normal GCs due to their lives at the centers of galaxies; this can result in extended star formation and the growth of massive black holes \citep[e.g.][]{Neumayer2020}.  Within the Milky Way, evidence for large metallicity and age spreads are seen in several luminous GCs, including $\omega$~Cen \citep[e.g.][]{Johnson2010,Villanova2014}, which appears to be the nucleus of a significant building block of the Milky Way \citep{Duncan2020,Pfeffer2021}, and M54 \citep{Siegel2007,Alfaro-Cuello2019}, which is the nucleus of the currently infalling Sagittarius dwarf galaxy.  An extended star formation history has also been seen in a massive (3$\times$10$^7$~M$_\odot$) UCD around NGC~4546 \citep{Norris2015}. 

It is now well established that supermassive black holes are common at the centers of galaxies with stellar mass $M_\star > 10^9$~M$_\odot$, including those with nuclear star clusters \citep{Miller2015,Nguyen2018,Nguyen2019,Greene2020}. Thus we expect stripped nuclei to host massive black holes as well. Supermassive black holes (SMBHs) have been dynamically detected in all five massive UCDs ($> \; 10^{7}$ ~M$_{\odot}$) with available high spatial resolution integral field spectroscopic data \citep{Seth2014m60,Ahn2017,Ahn2018,Afanasiev2018}. 
The black holes in these UCDs typically make up $\sim$10\% of their total stellar mass, similar to the mass fraction of the SMBHs in nuclear star clusters like that in the Milky Way \citep{Neumayer2020}.  
As noted above, simulations suggest that all UCDs in this mass range should be stripped nuclei \citep{Pfeffer2016,Mayes2021}.  These same simulations suggest that many stripped nuclei should have lower masses as well ($< \; 10^{7}$ ~M$_{\odot}$).  The star formation histories of $\omega$~Cen and M54 discussed above confirm that there are stripped nuclei at lower masses, as does the recent detection of a $\sim$10$^5$~M$_\odot$ black hole in Andromeda's most massive GC \citep{Pechetti2021}.  

There have also been many claims of detections of $\lesssim$10$^4$~M$_\odot$ black holes at the centers of Milky Way GCs \citep[see recent review by][]{Greene2020}.  However, none of these claims are regarded as robust \citep[e.g.][]{Zocchi2019,Henault-Brunet2020}, and no accretion evidence for any intermediate mass black holes in Milky Way GCs have been found \citep{Tremou2018}.  The claimed detections and upper limits for these black holes correspond to mass fraction of 0.1-1\% of the total mass of the cluster. Theoretically, there are processes that may form intermediate mass black holes during the formation and evolution of GCs, including the merging of massive stars early in the life of the cluster or the merging of stellar mass black holes over time \citep[e.g.][]{Portegies-Zwart2004,Inayoshi2020,DiCarlo2021} .  However, these mechanisms produce small black holes that make up $<$1\% of the total mass of the cluster,  thus the black holes found in UCDs are mostly far too massive to result from these processes without significant subsequent accretion that is unlikely outside a nuclear environment.  Therefore, the detection of high mass fraction black holes in luminous GCs provides strong evidence that they are stripped nuclei.  At the same time, as we discuss below, these high mass fraction black holes are not expected in all stripped nuclei.

Even before SMBHs were detected in UCDs, there was indirect evidence of their existence.  Central black holes raise the velocity dispersion of stars near the center of the stripped nuclei elevating the apparent dynamical mass-to-light ratio derived from integrated dispersion measurements relative to the expected mass-to-light ratio from stellar populations  \citep[e.g][]{Mieske2013,Duncan2014}. The enhanced mass-to-light ratio was used by \citet{Mieske2013} to predict the central black hole mass in UCDs with integrated dispersion measurements; they found similar mass fractions in many objects to the SMBH detections from resolved kinematics discussed above. More recently,
\citet{Karina2019} used the inflated mass-to-light ratio measurements to quantify the fraction of luminous GCs that have high mass fraction black holes (and thus are likely stripped nuclei) as a function of luminosity; they find that $\sim$20\% of luminous GCs at 10$^6$~L$_\odot$ have evidence for black holes, rising to $>$70\% at the highest luminosities.  This implies that a significant fraction of massive black holes in the local universe may be present outside of galaxy nuclei, a finding also shared by recent works \citep{Greene2020,Ricarte2021}. We expect a majority of these black holes in stripped nuclei to reside in lower luminosity nuclei than those that have currently been dynamically confirmed.  Dynamical detection of smaller black holes in Virgo is not possible with current technology, thus finding candidates of these lower mass stripped nuclei in nearby galaxies is important.



In this paper, we focus on finding additional candidate luminous GCs that may be stripped nuclei and have inflated mass-to-light ratios.   We use high-resolution spectroscopic data to derive internal velocity dispersion measurements and present size estimates for a new sample of luminous GC candidates around NGC~5128 (hereafter Cen~A). These luminous GCs extend out to 150 kpc in Cen~A's halo and are compiled from the candidates of \citet{KV2020} and \citet{Hughes2021}, making it the largest and most complete study of luminous GC velocity dispersions in the outskirts of Cen~A. Using the velocity dispersions and sizes, we estimate the dynamical mass-to-light ratio of each luminous GC. 

The paper is organized into six sections. In Section~\ref{sec:sample_selection_and_data_reduction} we describe the sample selection and observations. Section~\ref{sec:Vdisp Measurements} describes our method to obtain velocity dispersion and metallicities for our luminous GCs. In Section~\ref{sec:radii_lumin} we estimate radii and $V$-band luminosities, and then calculate mass-to-light ratios and black hole mass predictions in Section~\ref{sec:M/L}. Finally, we conclude in Section \ref{sec:conclusion}. Throughout this paper we apply a distance modulus for Cen~A of (m-M)$_{0}=$27.91 mag, corresponding to a distance of 3.8 Mpc.  We correct individual objects for foreground extinction, a typical Milky Way extinction value near Cen~A is A$_{V}=0.306$ mag \citep{Schlafly2011}.

\section{Sample Selection and Data Reduction}
\label{sec:sample_selection_and_data_reduction}
\subsection{Sample Selection and Sample Completeness} 
\label{subsec:sample_selection}
Based on the catalogs of \citet{KV2020} (hereafter KV20) and \citet{Hughes2021} we have targeted 314 GCs and stripped nuclei candidates around Cen~A's halo out to a projected distance of $\sim 150$~kpc$;\sim135\arcmin$. KV20 is a catalog of 614 luminous GC candidates in Cen~A based on {\em Gaia} DR2, and \citet{Hughes2021} is catalog of 40,502 GC candidates around Cen~A based on data from the Panoramic Imaging Survey of Centaurus and Sculptor (PISCeS) survey, combined with {\em Gaia} DR2 and NOAO source catalog data. Our targets span a much larger range of galactocentric distance in comparison to previous studies \citep[e.g][]{Rejkuba2007}.

In addition to the 314 targets from these catalogs, we have also selected two previously confirmed GCs in Cen A: \textit{aat329848} from \citet{Beasley2008} and \textit{vhh81-5} from \citet{Rejkuba2007} to provide us with some repeat measurements for testing the robustness of our results. We also targeted four nuclear star clusters (NSCs) of known Cen~A satellite galaxies (\textit{KK197-NSC}, \textit{ESO269-06}, \textit{Dw1\_NSC}, and \textit{DW3} from \citet{Denja2014,Denja2016}), and a diffuse cluster in Cen~A \textit{``Fluffy''}, that was found through visual inspection of PISCeS images, and whose radial velocity was presented in \citet{KV2020}. Thus in total, we observed 321 candidate luminous GCs. Of these 321 objects, we were able to measure the radial velocity for 219 candidates, of which 165 also had measurable velocity dispersions (see \S\ref{subsec:spectral_fitting}). Based on the radial velocities, we identify 78 as being Cen~A clusters, and we measure the velocity dispersion and mass-to-light ratios (\S\ref{sec:M/L}) for 57 of these.

To quantify the completeness of our sample of luminous GCs around Cen~A, we first look at the total sample of luminous GCs around Cen~A in the literature.  We use the compilation of velocity and {\em Hubble} confirmed GCs in \citet{Hughes2021} and make a magnitude cut of  $g\leq19.1$ to define a clean sample of luminous GCs. This cut corresponds to $g_0 = 18.8$, and $M_V \sim -9.5$ ($L_V \sim 5\times10^{5}$ L$_\odot$) and is the same cut used by KV20. We also confirm here 13 clusters from KV20 as Cen~A GCs based on their radial velocities.  Thus in total there are 118 clusters with $g\leq19.1$ with confirmed radial velocities in the literature.  Of these 118 GCs with $g\leq19.1$, 27 have previous reliable velocity dispersion measurements  \citep{Harris2002,martini2004,Rejkuba2007}, while in this work we present 36 new measurements as well as remeasuring dispersions in 10 objects. Thus in total, 63 of the 118 GCs with $g\leq19.1$ now have reliable velocity dispersion measurements.  We also measure dispersions for 9 Cen~A GCs and 2 additional nuclear star clusters at $g > 19.1$.  Also, in the calculation above we exclude the large number of dispersion measurements from \citet{Taylor2010} and \citet{Taylor2015}, as the median signal-to-noise of their measurements is very low \citep[2 for][]{Taylor2015}, well below the limit where we find we can reliably measure dispersions using similar high-resolution spectroscopy.  Just a handful of their measurements have signal-to-noise in the range of the measurements we present here.  Issues with both their published velocities and dispersions have been documented previously \citep{Karina2018,Karina2019}, and we find additional issues that we identify below.

\begin{figure}
\plotone{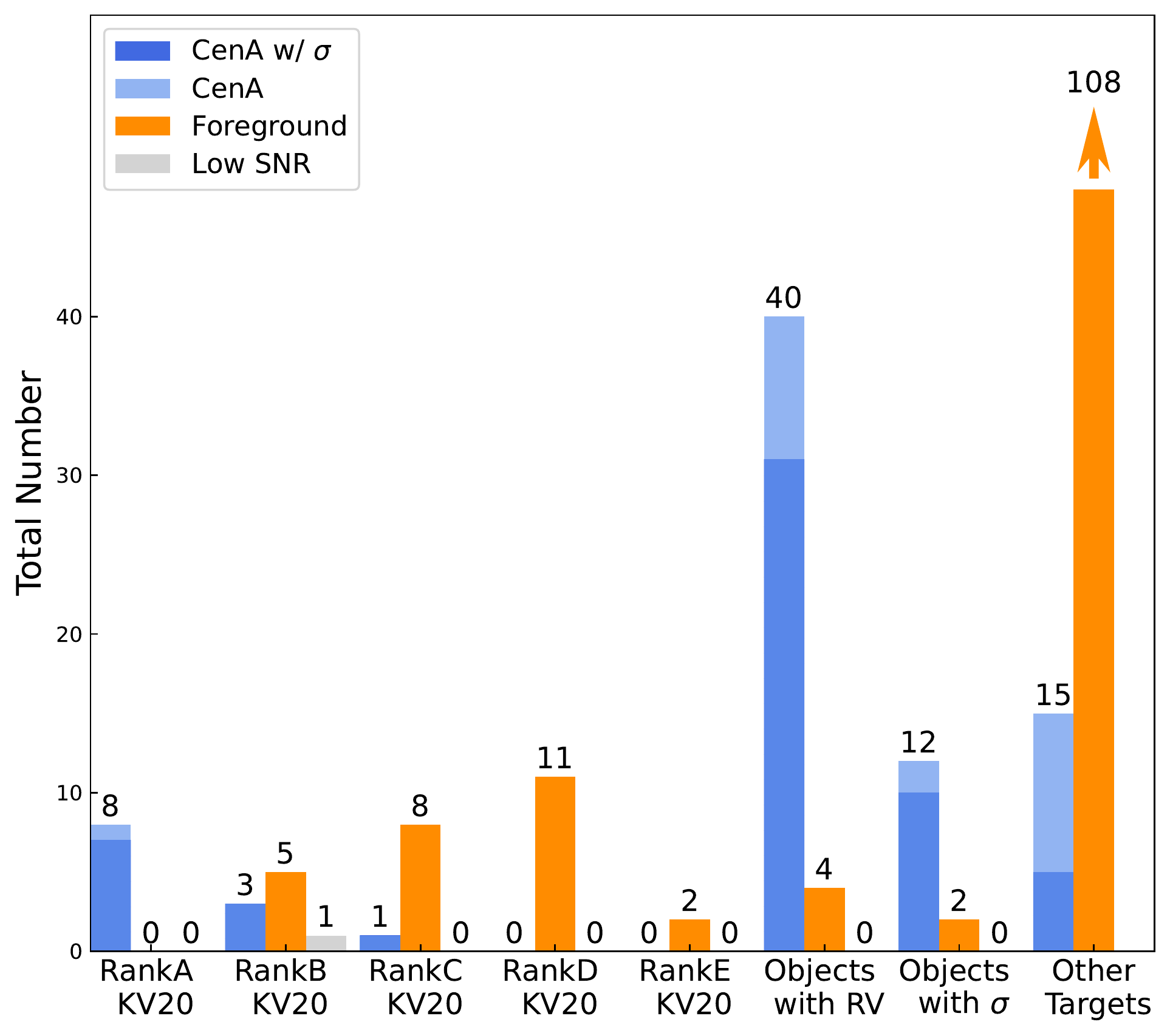}
\caption{Histogram of 321 luminous GCs candidates in our sample categorized as ”Cen~A Objects” and ”Non Cen~A Objects” based on their radial velocities.  We consider objects with RV$\geq$250 km/s to be Cen~A objects, while those with RV$\leq$250 km/s respectively are likely Milky Way foreground stars (see \S\ref{subsec:spectral_fitting} for details).  RankA-E KV20 correspond to objects from the \citet{KV2020} catalog, with the A ranked objects being the highest likelihood to be GCs in Cen~A. We also show objects with previous radial velocity and/or velocity dispersion ($\sigma$) measurements.   
\label{fig:Hist}}
\end{figure}

We can also estimate the total number of luminous GCs in Cen~A based on our confirmation of KV20 candidates.  The KV20 sample is a nearly complete catalog of new luminous GC candidates at galactocentric radii beyond 5.5~kpc. Fig.~\ref{fig:Hist} shows the results of how many of our observed targets are confirmed as Cen~A clusters based on their radial velocities (see \S\ref{subsec:spectral_fitting}). Using these data for the different ranks of cluster candidates in KV20, we estimate there are a total of $\sim$31 real Cen~A GCs in the catalog, of which we have found 13 so far.  Combined with previously identified clusters that likely form a complete sample at smaller radii, we estimate a total of $\sim$136 total luminous GCs with $g \leq 19.1$ around Cen~A.


Throughout this paper, we use photometry for our sources from the NOAO DR2 source catalog \citep{noao_dr2}. This photometry in ugriz is derived from DECam observations, and appears to be more reliably calibrated and more complete than photomety from the same data presented in \citet{Taylor2017}. Throughout this paper, we present magnitudes corrected for galactic extinction using \citet{Schlafly2011}.  

\subsection{Spectroscopic Data of Luminous GC Candidates} \label{subsec:spectroscopic_data}

The observations of 321 GC candidates in Cen~A were taken over several observing runs between February 2017 and March 2019 using the Michigan/Magellan Fiber System (M2FS) and the Magellan Inamori Kyocera Echelle (MIKE) spectrograph at the $6.5$~m  Magellan Clay telescope at Las Campanas Observatory, Chile.  The list of M2FS plates and nights of MIKE observations are shown in Table \ref{table:Observations}. 

M2FS is a fiber-fed double spectrograph optimized to operate from $3700-9500$~\AA \citep{M2FS}. Each spectrograph is fed by 128 fibers with a diameter of $1.2\arcsec$. M2FS's large field of view (30\arcmin~diameter) and its high spectral resolution  (R~$\simeq \: 25000 ; \: \sim 10$~km/s) are a perfect combination to resolve the velocity dispersions of multiple GCs candidates simultaneously.  

MIKE is a single slit echelle, double arm spectrograph, delivering full wavelength coverage from $3350-5000$~\AA~(blue) and $4900-9500$~\AA~(red) \citep{Mikeref}. We used a $1$~\arcsec~ slit width to obtain a spectral resolution of R~$\simeq \: 22000; \: \sim 13$~km/s across the full wavelength range. 

Our observation plan was to use M2FS for regions around Cen~A where three or more high priority luminous GC candidates (see below) are observable simultaneously, and target the rest of the high-priority candidates individually with MIKE. For targets observed with M2FS we used two distinct setups.  First we observed using the MG\_Wide filter, providing 4 orders ranging from $5119-5443$~\AA~(hereafter Magnesium Triplet). We focused our analysis of these spectra on the second order spanning $5133-5218$~\AA~ where the magnesium triplet lines ($5167~\AA$, $5172~\AA$, $5183~\AA$) are located. The second setup uses the $CaIRT_{041}$ filter covering a single order $8471-8819$~\AA, which includes the "Calcium Triplet" lines at  $8498~\AA$, $8542~\AA$, and $8662~\AA$
The Calcium Triplet setup provides higher signal-to-noise (S/N) relative to the Magnesium Triplet but with stronger sky lines.  These data allowed us to derive dispersions for fainter candidates.   
The MIKE data simultaneously provide full wavelength coverage from $3350-9500$ \AA. 
Our kinematic analysis of the MIKE data focused on the same regions studied with M2FS; specifically, the Magnesium Triplet order covering $5140-5260$~\AA,~and for the Calcium Triplet we use the order covering $8461-8708$~\AA.

The seeing for the observations was determined from the Las Campanas DIMM seeing monitor measurements during M2FS observations, and from direct measurements for the MIKE observations.  The seeing is given in Table~\ref{table:Observations}.  

\begin{deluxetable}{l|llllr}[h!]
\tabletypesize{\scriptsize}
\caption{Observation log}
\label{table:Observations}
\fontsize{6}{6}\selectfont
\tablecolumns{6}
\tablewidth{0.5\columnwidth} 
\tablehead{ \colhead{Observation} & \colhead{Instr.} & \colhead{Setup} &\colhead{ExpTime} & \colhead{Seeing} & \colhead{Obs Date}  \\
\colhead{} & \colhead{}  & \colhead{} & \colhead{[min]} & \colhead{["]} & \colhead{mm/dd/yyyy} }
\startdata
        FieldMG 0 &  M2FS &  Mg &  4$\times$45 & 0.65 &02/24/2017\\
        FieldMG 1 &  M2FS &  Mg &  4$\times$45 & 0.9  &05/20/2017\\
        FieldMG 2 &  M2FS & Mg  &  4$\times$45 & 1.2  &05/21/2017\\
        FieldMG 3 &  M2FS & Mg  &  4$\times$40  & 0.77 &05/30/2017\\
        FieldMG 4 &  M2FS &  Mg &  4$\times$48 & 0.8 &06/03/2017\\
        FieldMG 6 &  M2FS & Mg  &  3$\times$50 & 1.1 &05/11/2018\\
        FieldCa 1 &  M2FS & Ca  &  3$\times$40 & 0.7   &02/26/2019\\
        FieldCa 2 &  M2FS & Ca  &  3$\times$67 & 0.7   & 03/01/2019\\
        FieldCa 3 &  M2FS & Ca  &  3$\times$34 &  0.7  & 03/04/2019\\
        Night 1   & MIKE & Full &  180 & 0.85-1.02 & 06/17/2018\\
        Night 2   & MIKE & Full &  270 & 0.7-1.2 & 04/05/2019\\
        Night 3   & MIKE & Full &  405 & 0.8-1 & 04/06/2019 \\
\enddata
\tablecomments{Mg and Ca in column 3 refers to M2FS Magnesium and Calcium Triplet setup respectively. MIKE delivers full wavelength range coverage that includes Magnesium and Calcium Triplet and is denoted here as ``Full'' (See \S\ref{subsec:spectroscopic_data}). Column 4 shows the number of exposures times the exposure time. For MIKE observations each individual objects hasa different exposure time, and we only show the combined exposure time. }
\end{deluxetable}

\subsection{M2FS Data Reduction \label{subsec:m2fs_reduction}}

The data reduction for M2FS was done using a combination of tasks in IRAF following the manual by Christian I. Johnson \footnote{\url{https://www.cfa.harvard.edu/oir/m2fsreduction.pdf}}. 

The data reduction of the science frames requires a set of flat-fields, Thorium-Argon lamps, and bias frames for calibration. Each image observed with the CCD camera in M2FS is read out through four amplifiers that splits the image into four frames. Each amplifier frame needs to be trimmed to remove the overscan region, bias subtracted, and rotated to be correctly oriented to join them into a single frame.  We then remove cosmic rays from the combined images using the IRAF routine \textit{L.A cosmic}.

At this point the images are ready for data reduction with {\sc dohydra}. {\sc dohydra} is a sophisticated script optimized for the reduction of data from the WIYN and the \textit{Blanco Hydra} spectrographs, but it can also be used for the data reduction of any similar multi-fiber spectrograph such as the M2FS. The first step in {\sc dohydra} is the fiber identification using the luminous twilight spectrum in the flat-fields. For the Magnesium Triplet setup, each individual fiber has four orders on the detector, thus only 32 objects per plate are observable.  For the Calcium Triplet setup each fiber has only a single order, enabling observations of 128 separate objects per plate.
For the Calcium Triplet we trace every fiber, as each fiber correspond to a single target. For the Magnesium triplet we trace only the second order, as this corresponds to the wavelength range with the Magnesium Triplet lines. 

Typically flat-fielding would be performed next, however we found that flat fielding resulted in noisier spectra. For that reason we did not correct for the flat fields, as accurate flux calibration is not important for our kinematic measurements.  Finally, wavelength calibration was done using a Thorium-Argon (ThAr) lamp atlas \citep{ThoriumArgon_Atlas} to identify lines in the lamp frames. The typical RMS of our wavelength solution was about a third of a pixel; the pixel size in the Calcium Triplet order is 0.11~\AA, while for Magnesium Triplet it is 0.06~\AA. For the Calcium order data, the strong sky lines enabled us to check our wavelength solution -- we find a standard deviation of the line wavelengths relative to expected of $0.076$~\AA~ suggesting an absolute velocity errors of $<3$~km/s.  No sky subtraction was performed (instead we fit the sky during our kinematic fits, see Section \ref{subsec:spectral_fitting}).

After the data reduction with {\sc dohydra} was completed, we summed the spectra from multiple exposures into a single spectrum in python for each object to increase the S/N. Additionally, {\sc dohydra} returns an error spectrum for each exposure. We summed in quadrature the error spectra of multiple exposures to create a single error spectrum.

A total of 301 targets were observed with M2FS, of them 116 in the Magnesium Triplet, and 185 in the Calcium Triplet setup. Our targets are brighter in the Calcium Triplet, providing significantly higher S/N than the Magnesium Triplet order. The 301 M2FS targets span a range of $g$ absolute magnitude of $-12.8$ to $-5.3$ mag, with a median of -9.5. Figure~\ref{fig:SNR_vs_gmag} shows the S/N ratio and $g$ mag brightness for all our luminous GC candidates. For the five luminous GCs with large discrepancies between the NOAO DR2 and {\em Gaia} photometry (see \S\ref{app:magnitudes}), we present estimated $g$-band magnitudes based on the median $g-V$ color ($g-V = 0.18$) of the rest of our luminous GC candidates.

\subsection{MIKE Data Reduction}\label{subsec:mike_reduction}
We observed 20 luminous GC candidates with MIKE, along with 22 known foreground stars on two nights from the 5-7. April 2019 and on one night on  2018 June 16 and 17. 
The spectra were reduced with the CarPy reduction pipeline \citep{Kelson2000, Kelson2003}. The pipeline automatically applies all the standard steps of applying the bias and flatfield to the data. In addition it directly takes care of the sky subtraction as well as the rectification of the distorted spectra and cosmic ray removal. 

\begin{figure}[H]
\plotone{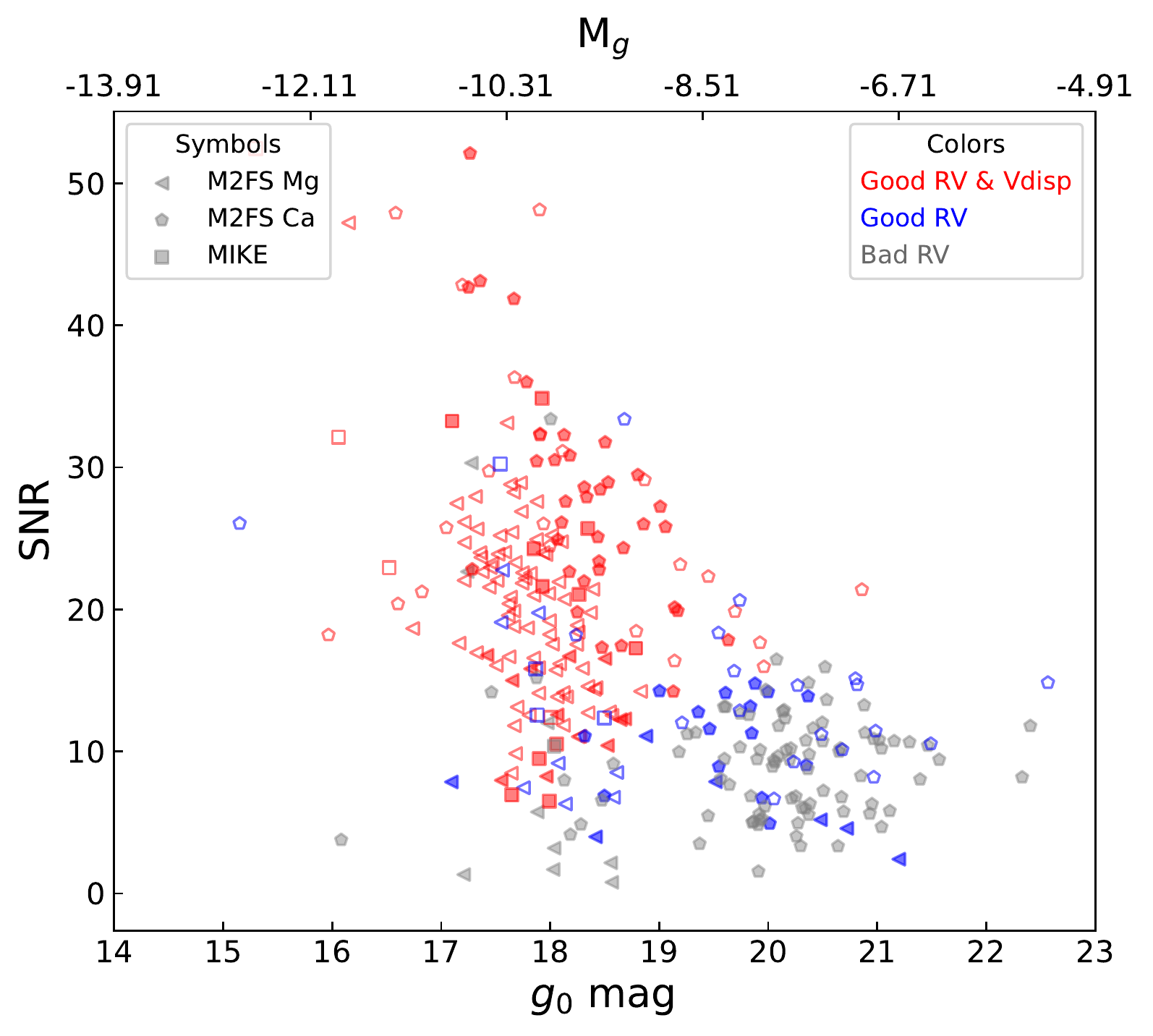}
\caption{ Summary of S/N ratios and NOAO DR2 $g$ magnitude for the 321 observed luminous GC candidates. We measure the S/N as the median of the S/N per pixel across the full wavelength range observed. We identify candidates observed with M2FS in Magnesium Triplet with triangular markers, Calcium Triplet with pentagon markers, and targets observed with MIKE Calcium Triplet with square markers. Red objects represent targets that have visually verified radial velocities and velocity dispersions, while blue objects only have visually verified radial velocities. Gray objects are targets for which we cannot obtain kinematic measurements. Solid and open markers represent targets with radial velocity consistent with Cen~A or foreground stars respectively.
\label{fig:SNR_vs_gmag}}
\end{figure}

The S/N for the MIKE Calcium Triplet data as a function of $g$-band magnitude is shown in squares in Figure~\ref{fig:SNR_vs_gmag}. The 20 MIKE targets span a range of $g$ absolute magnitude of $-12.6$ to $-9.2$ mag. Additionally, we had access to the fully reduced MIKE spectra of 14 luminous Cen~A globular clusters from \citet{martini2004} used for testing our spectral fits (see \S\ref{subsec:spectral_fitting})

\section{Kinematic \& Metallicity Measurements}\label{sec:Vdisp Measurements}

\subsection{Template Selection}
We fitted the spectra of every candidate observed with M2FS and MIKE using the Penalized Pixel-Fitting pPXF routine \citep{ppxf}. pPXF requires a set of templates to model the spectra. We used the high resolution (R$=500000$) synthetic stellar templates from the Phoenix library \citep{phoenixlib}. We chose the temperature and log(g) of our Phoenix stellar templates based on the Padova Isochrones \citep{PARSEC-COLIBRI} with an age of $10$~Gyr, selecting stars along the subgiant, giant, and horizontal branch as these are expected to dominate the light of globular clusters. We selected templates with metallicities ranging from [Fe/H] =  0.0 to [Fe/H] = -2.0 with an interval of 0.5 dex, and log(g) from 0 to 4 with an interval of 0.5 dex. The temperatures of the templates ranges from 3200 -- 11200~K.  We note that we are not assuming a $10$~Gyr age for our templates, but rather use a $10$~Gyr isochrone to select the approximate log(g), metallicity and temperature of the appropriate Phoenix templates since these properties determine the shape of the spectral lines of the templates.
This resulted in a library of 46 total stellar templates.

\subsection{LSF determination}

The synthetic stellar templates need to be convolved with the intrinsic line-spread function (LSF) of the observed spectra before fitting.
We derive the LSF by fitting individual stars observed in each setup with our Phoenix templates using pPXF. We observed stars of known spectral type during the observing runs with MIKE, and those are used to determine the LSF. For M2FS, we did not observe stars of known spectral type, but many of our luminous candidate GCs were found to be Milky Way foreground stars based on their radial velocities  ($V < 250$~km/s). Assuming the spectrographs have Gaussian LSFs, the derived velocity dispersion for the stars measures the width of the intrinsic LSF of the instrument. The measured dispersion can be converted into FWHM (in Angstroms) $FHWM = 2 \sqrt{2 ln(2)} \sigma_{star}\lambda_{c}/c$; with $\sigma_{star}$ being the derived velocity dispersion, $c$ the speed of light, and $\lambda_{c}$ the central wavelength in each order. We used the median FWHM for all the stars with S/N higher than 20 as the LSF (see Figure \ref{fig:LSF_lot} in the Appendix). We obtained a different LSF for each filter (Calcium and Magnesium Triplet) and instrument. The error of the LSF was estimated by taking the standard deviation of the FWHM values, and this was propagated into the dispersion measurements later. The velocity dispersions measured in our luminous GC candidates are much higher than the width of the Gaussian LSF; therefore, the uncertainty in the LSF and errors caused by any non-Gaussianity of the LSF do not significantly impact our results.  



\subsection{Spectral fitting}\label{subsec:spectral_fitting}
To ensure that our globular clusters are well fit by our selected templates, we optimized our pPXF parameters and template selection to maximize consistency with the velocity dispersion measurements of the 14 massive globular clusters from \citet{martini2004}. Specifically, we removed two of our initial templates and set pPXF's multiplicative polynomial degree to 12.
No additive polynomial was used to preserve relative line depths. Our pPXF fits were sensitive to the initial radial velocity guess, with objects with bad initial guesses often resulting in bad fits with unrealistically large dispersions.  The initial radial velocity guess was obtained for each luminous GC candidate by running a grid of different radial velocities guesses from $0-900$ km/s spaced by $100$ km/s, and keeping the results from the fit with the best reduced $\chi^{2}$. Once a radial velocity guess was obtained a bad-pixel mask was created by performing a 2-sigma clipping to remove outliers. 

\begin{figure*}
\gridline{\fig{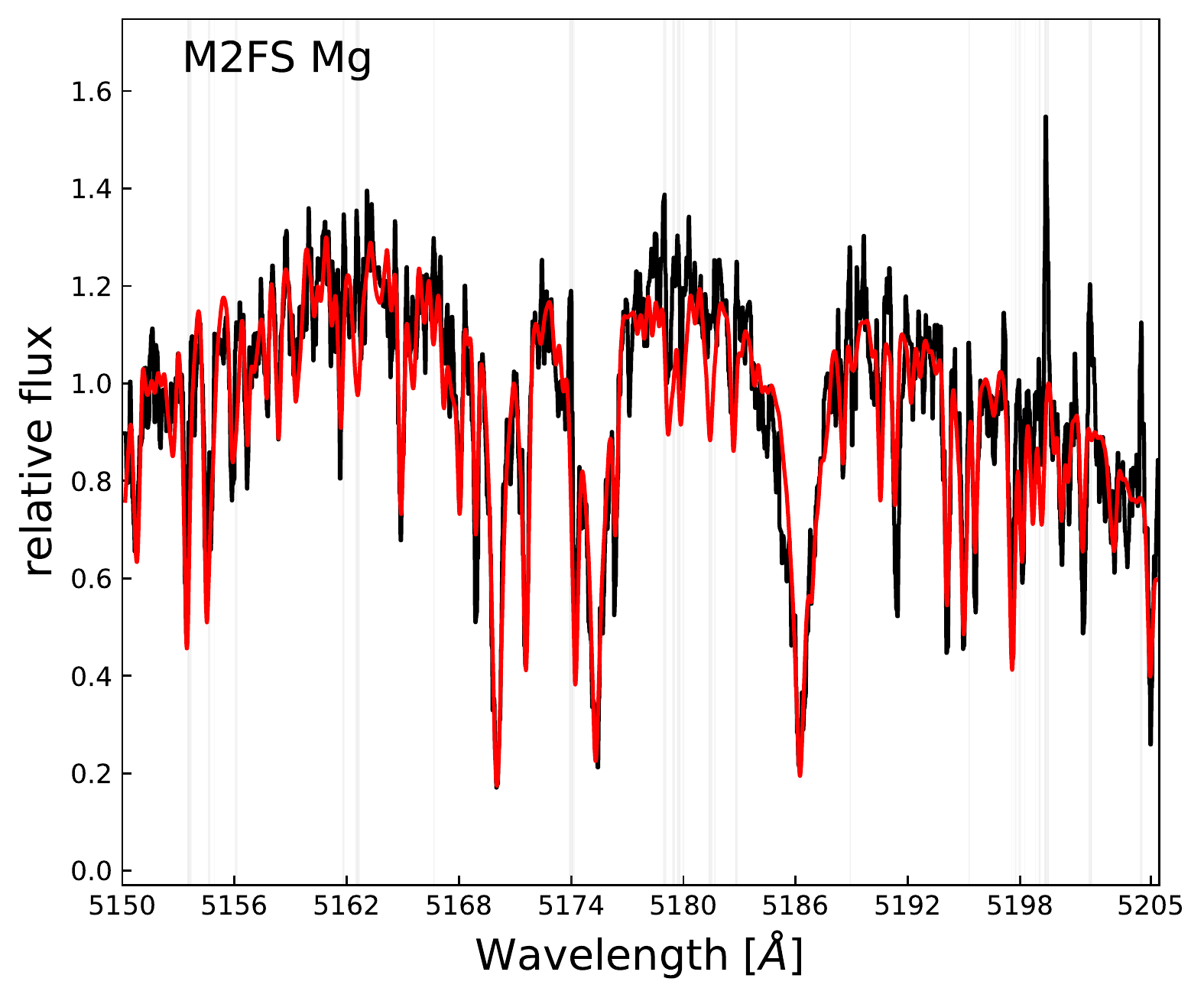}{0.33\textwidth}{(a)}
          \fig{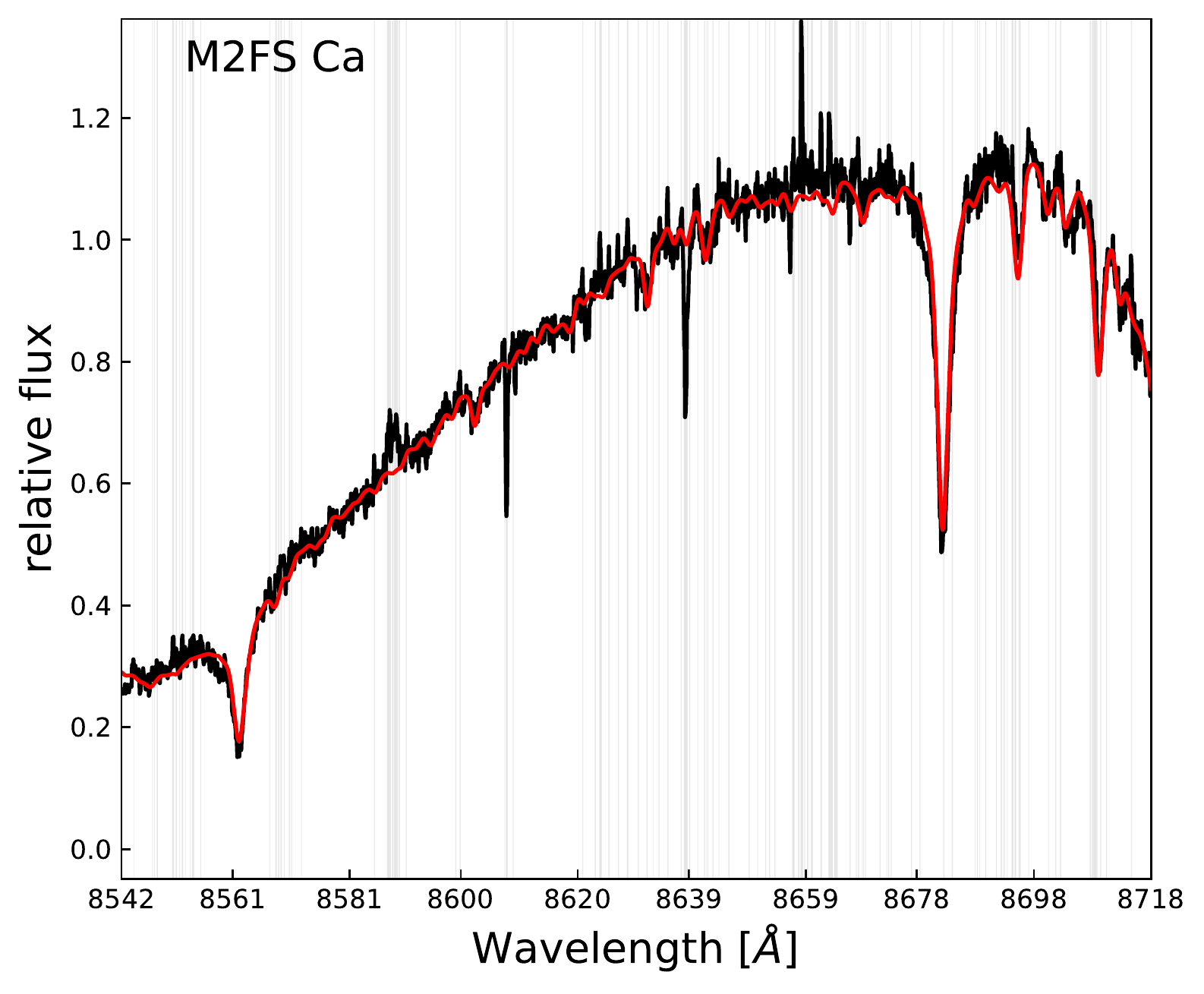}{0.33\textwidth}{(b)}
          \fig{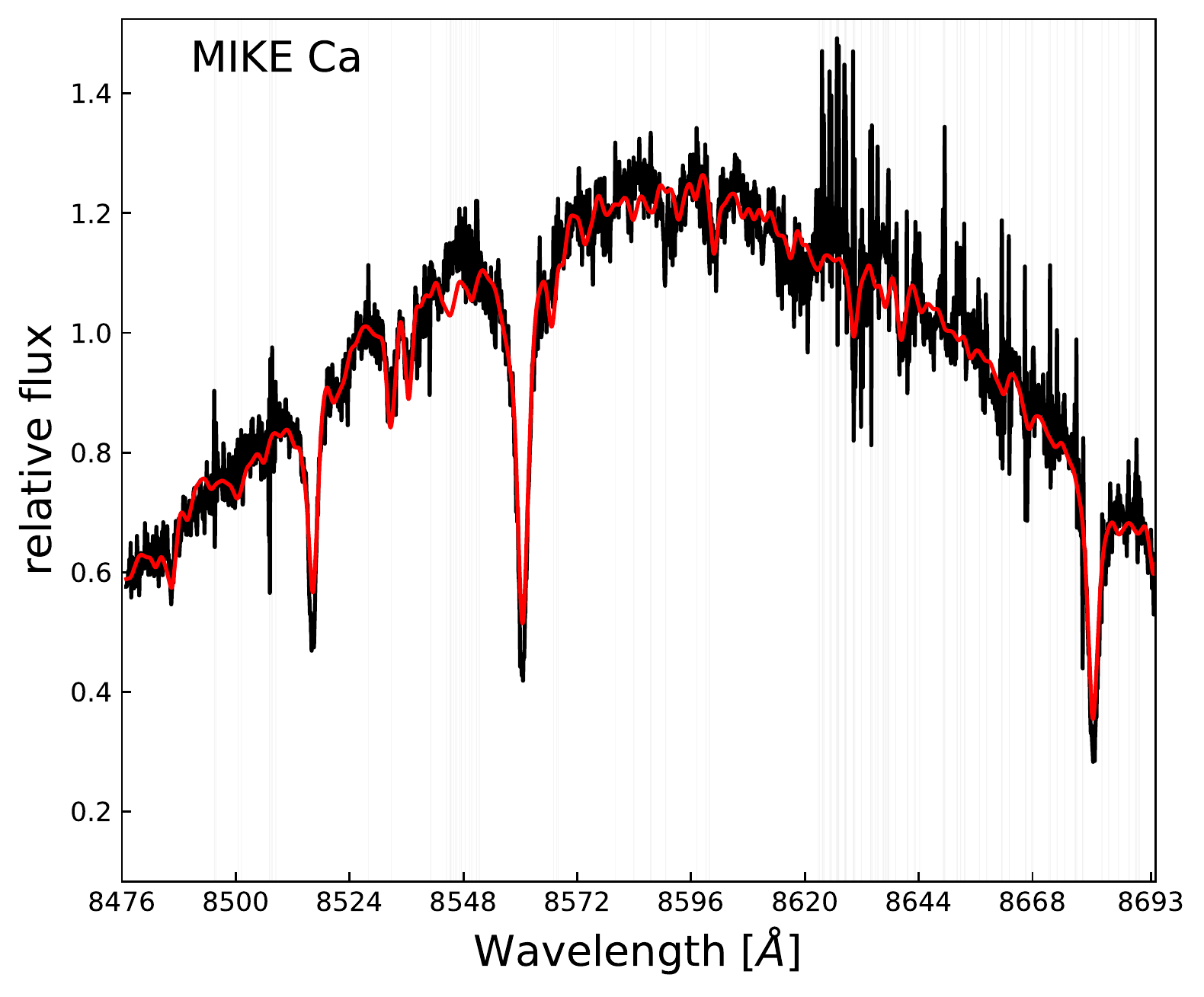}{0.33\textwidth}{(c)}
          }
\caption{Spectral fit using pPXF examples for three luminous GCs in the M2FS Magnesium Triplet (M2FS Mg) (a), M2FS Calcium Triplet (M2FS Ca) (b), and MIKE Calcium Triplet (MIKE Ca) (c).  The  y-axis represents the relative flux per unit of wavelength.  The observed spectrum (black) is fitted using the combination of stellar models (red). Gray vertical lines show masked regions include sky emission lines.\label{fig:example_fit} }
\end{figure*}

Since no sky subtraction was done during the data reduction for the M2FS data, for the Calcium Triplet M2FS data, we fit the sky emission lines with spectra from dedicated sky fibers in our field-of-view. After an initial fit, we examined the residuals at the wavelengths of known sky lines \citep{Hanuschik2003}, and masked any pixels where the sky subtraction left significant residuals. For the MIKE data, sky subtraction along the slit was done during the data reduction; this also left behind sky residuals in some cases, thus we apply the same masking procedure to the Calcium Triplet MIKE data. Our final radial velocity and velocity dispersion measurement pPXF runs use these initial radial velocity guesses and bad pixel masks as input.  An example of the fits for each setup is shown in Figure \ref{fig:example_fit}. 

To estimate our errors on the radial velocity and dispersion measurements, we first re-calculated the radial velocity and velocity dispersion using the lower and upper LSF values (see Figure~\ref{fig:LSF_lot}), and then performed Monte Carlo simulations based on the error spectra for each object. Specifically, we added noise to the observed spectrum in every iteration based on its error spectrum and re-fit in pPXF. This was repeated 50 times for each object.

\begin{figure*}
\gridline{\fig{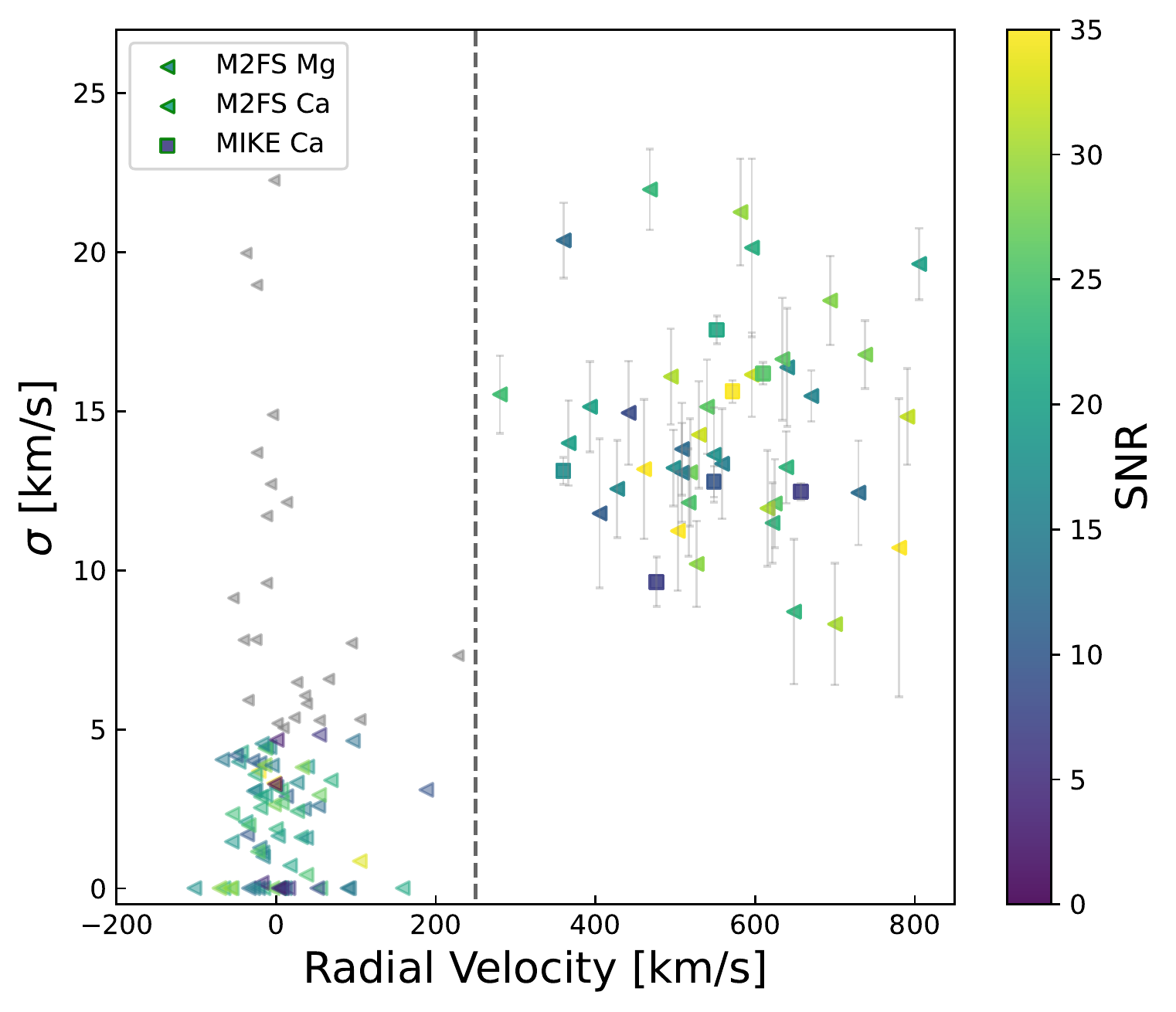}{0.5\textwidth}{}
        \fig{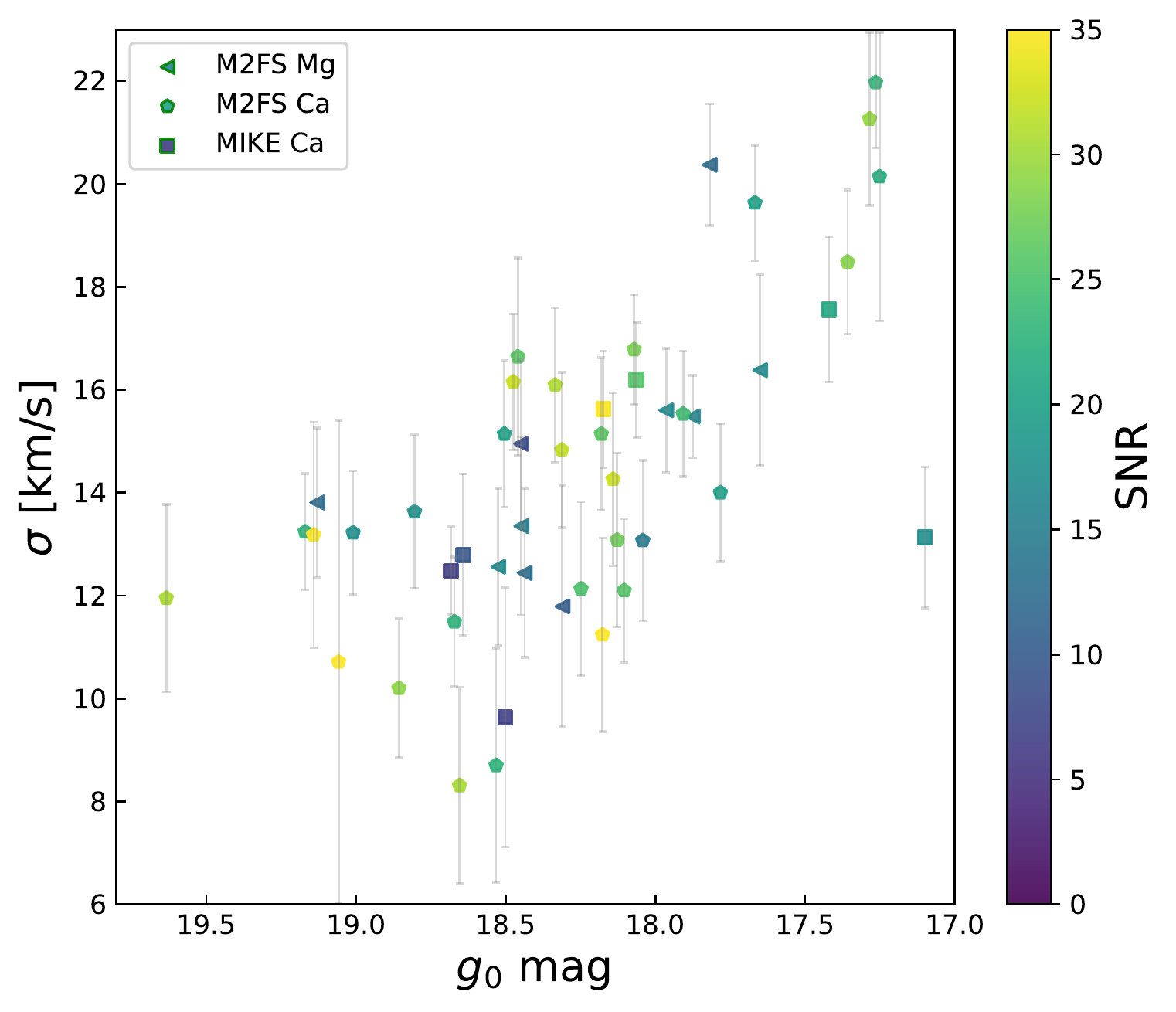}{0.5\textwidth}{}\label{fig:vdisp_gmag_plot}}

\caption{\textit{Left:} Radial velocity vs. velocity dispersion for 165 luminous GC candidates with measurable dispersions. The black vertical dashed line shows the minimum radial velocity for Cen~A GCs ($250$ km/s). Gray symbols show likely binary stars.
 \textit{Right:}  Relation between $g$ mag and velocity dispersion for the 57 luminous GCs with visually verified velocity dispersion. We expect a relation between these two quantities assuming that mass follows light, with brighter GC being more massive, and a $g$ mag $=18.7$ corresponds roughly to a luminous GC mass of $10^{6}$M$_{\odot}$ assuming a mass-to-light ratio of 2.\label{fig:Rv_vs_sigma}}
\end{figure*}

Figure~\ref{fig:Rv_vs_sigma} summarizes the radial velocity and velocity dispersion measurements for the 165 luminous GC candidates with measurable velocity dispersions. Based on previous spectroscopic studies \citep[e.g][]{Peng2004}, we use a radial velocity cut of 250~km/s to separate Cen~A objects from possible foreground objects. Most targets with radial velocities lower than 250~km/s have significant ($>$2$\sigma$) {\em Gaia} EDR3 proper motions, and thus they are consistent with being Milky Way foreground stars. The target with radial velocity $\sim$230~km/s (\textit{H12\_82}) has a significant {\em Gaia} EDR3 proper motion, suggesting it is a foreground star.  
High velocity dispersion foreground objects are most likely binary stars; visual inspection shows double lines in many cases. None of the Cen~A objects showed similar double line profiles.

\subsection{Comparison to Literature \& Reliability Assessment}\label{subsec:comp_to_lit}

\begin{figure}
\plotone{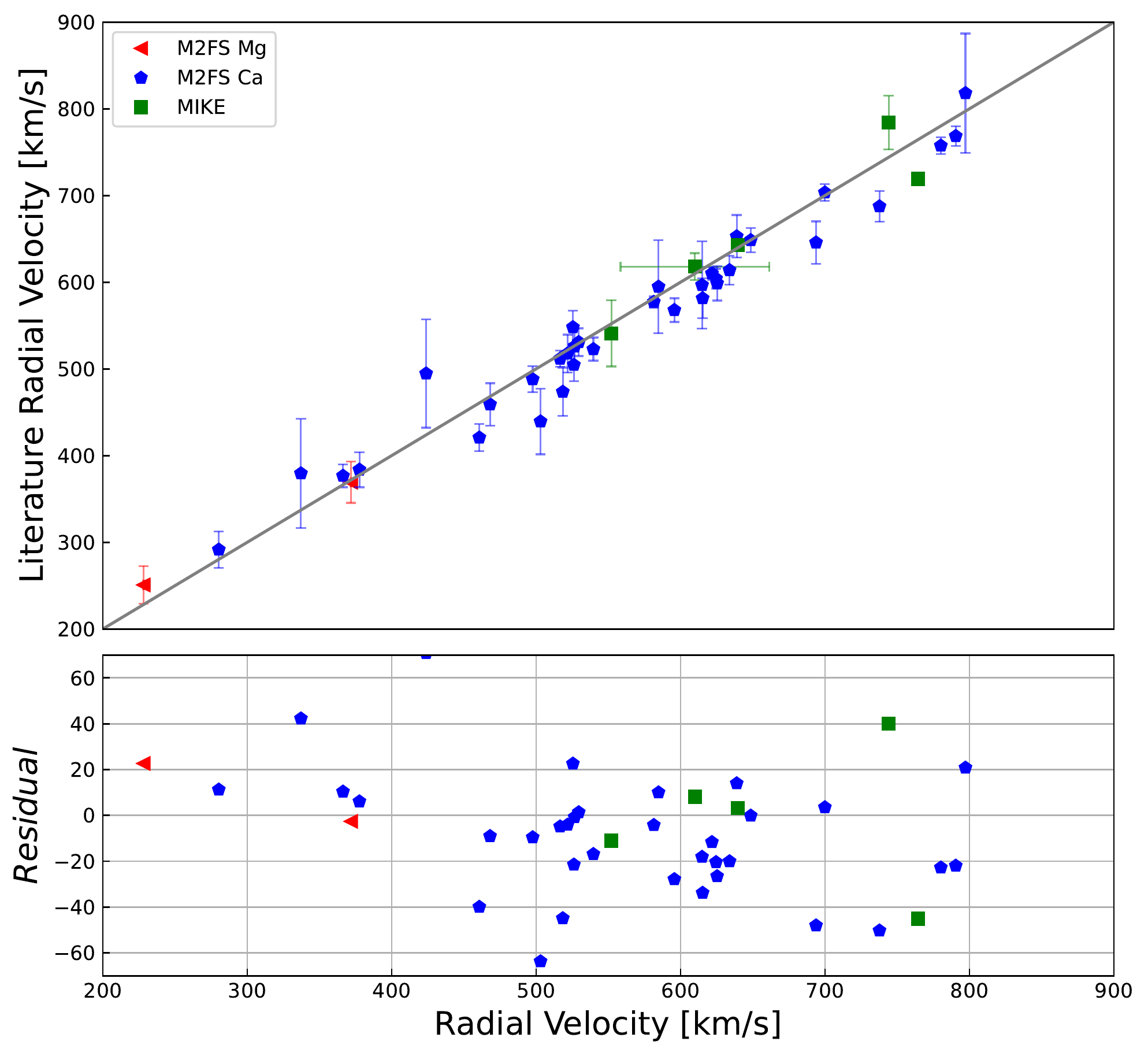}
\caption{ Comparison of 41 radial velocities in common between this work and  literature values from the literature \citep{Peng2004,Woodley2005,Rejkuba2007,Beasley2008,Woodley2010}. The solid black line shows the one-to-one relation. Symbols represent the different configurations as described in Figure~\ref{fig:Rv_vs_sigma}. The standard deviation of the residuals is 20~km/s.
\label{fig:RV_lit}}
\end{figure}

Seventy of our luminous GC targets have literature radial velocity measurements  \citep{Peng2004,Woodley2005,Rejkuba2007,Beasley2008,Woodley2010}. Out of these 70 targets, 13 have available velocity dispersions. Figure~\ref{fig:RV_lit} shows the comparison of radial velocity measurements with literature values.  We find a good agreement between the two with a scatter of $20$~km/s and minimal bias. This scatter is in excess of reported velocity measurements in many cases. However, we think that this is a result of errors being underestimated in previous work; \citet{Beasley2008} found a 35 km/s offset and 60 km/s scatter in comparing his work to that of \citet{Peng2004}. We reclassify four targets as foreground stars that were previously classified as Cen~A GCs based on their radial velocities: these include two targets (\textit{HH-32} and \textit{AAT112964}) from \citet{Beasley2008} and one object from (\textit{T17-1974}) \citet{Taylor2015}\footnote{The previous velocities for these objects are: 480$\pm$27.4 for \textit{HH-32}, 439$\pm$97.9 km/s for \textit{AAT112964} and 503$\pm$8 km/s \textit{T17-1974}; our derived velocities are listed in Table~\ref{tab:all_GC_data}.}.

\begin{figure}
\plotone{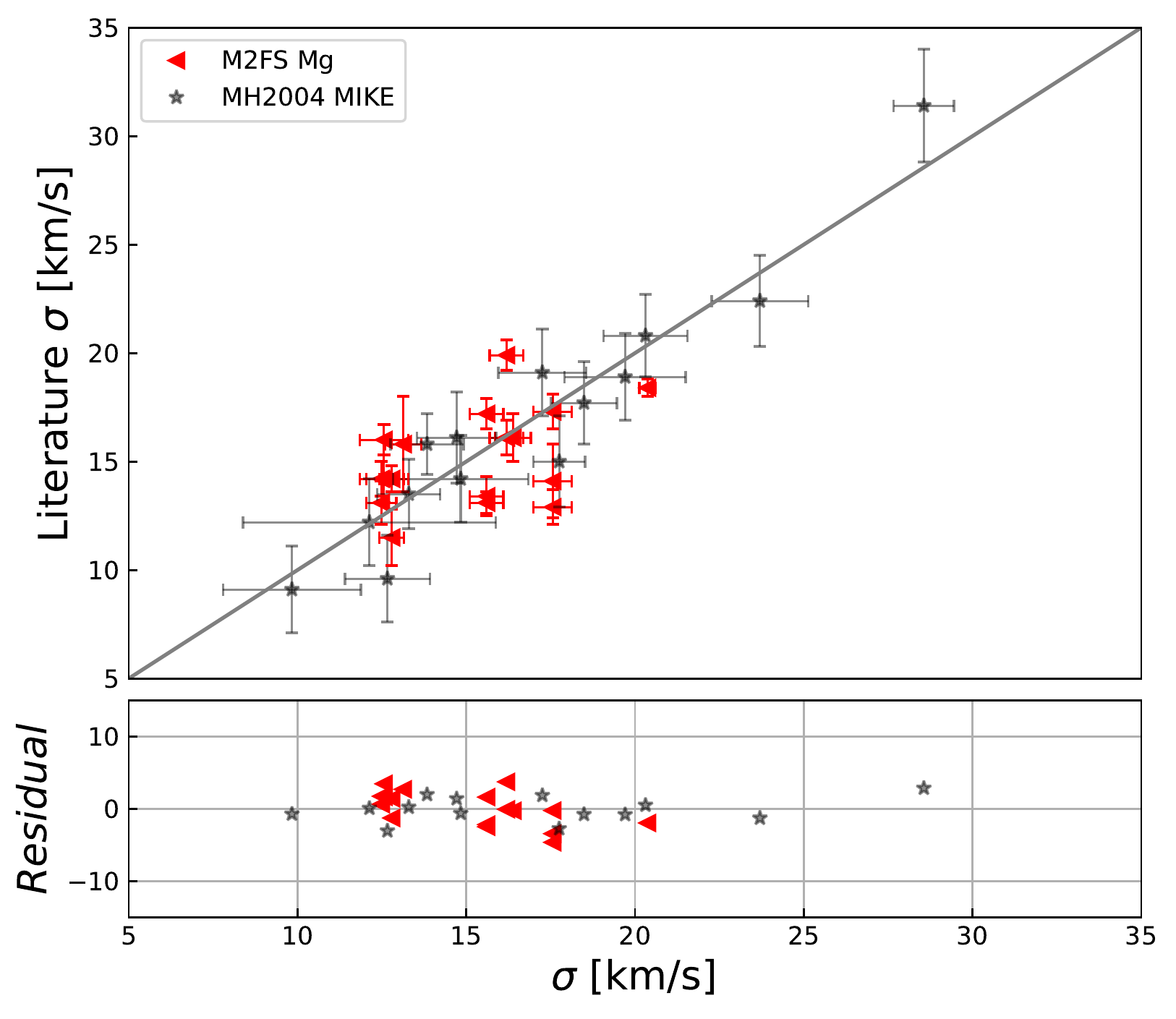}
\caption{ Comparison of 14 velocity dispersion estimates in common between this work and literature literature values from \citet{Harris2002,martini2004,Rejkuba2007,Taylor2010,Taylor2015}. The MH2004 MIKE data show the comparison to our analysis of the \citet{martini2004} data compared to their published values. The standard deviation of the residuals is 2~km/s. 
\label{fig:Vdisp_lit}}
\end{figure}

For the thirteen luminous GCs with previous velocity dispersion measurements from \citet{Rejkuba2007}, we can determine the velocity dispersion for nine of them. Their comparison with literature values is shown in Figure~\ref{fig:Vdisp_lit}. The velocity dispersions of the 14 luminous GC from \citet{martini2004} used to set-up our pPXF fits (see \S\ref{subsec:spectral_fitting}) are shown in gray in Figure~\ref{fig:Vdisp_lit} as they are not part of the new measurements of luminous GCs presented here.  We find a very good agreement with literature values, with a standard deviation of the residuals of $2$~km/s which are consistent with our velocity dispersion uncertainties. We do not include the S/N $<5$ velocity dispersion measurements for \textit{T17-1974} in \citet{Taylor2015} found to have a Milky Way foreground velocity.

The reliability of our dispersion measurements clearly correlates with S/N. 
However,  due to varying positions of sky lines, and outlier pixels in the spectra, a simple S/N cut did not cleanly separate good and bad measurements. For that reason, we visually inspected the shape and depth of multiple absorption lines in the fit for each luminous GC candidate to determine if the fit was reliable; the kinematic fits passing this inspection are hereafter referred to as ''visually verified'' measurements.
Fig.~\ref{fig:SNR_vs_gmag} shows that our final visually verified dispersions are clearly correlated with S/N, with most spectra above S/N of 10 providing reliable dispersion measurements.  We also find that almost all clusters with visually verified dispersions have $g_0 \lesssim 19$.



\subsection{Repeated Measurements and Uncertainties}
\label{subsec:repeated_meassurements}
\begin{figure}
\plotone{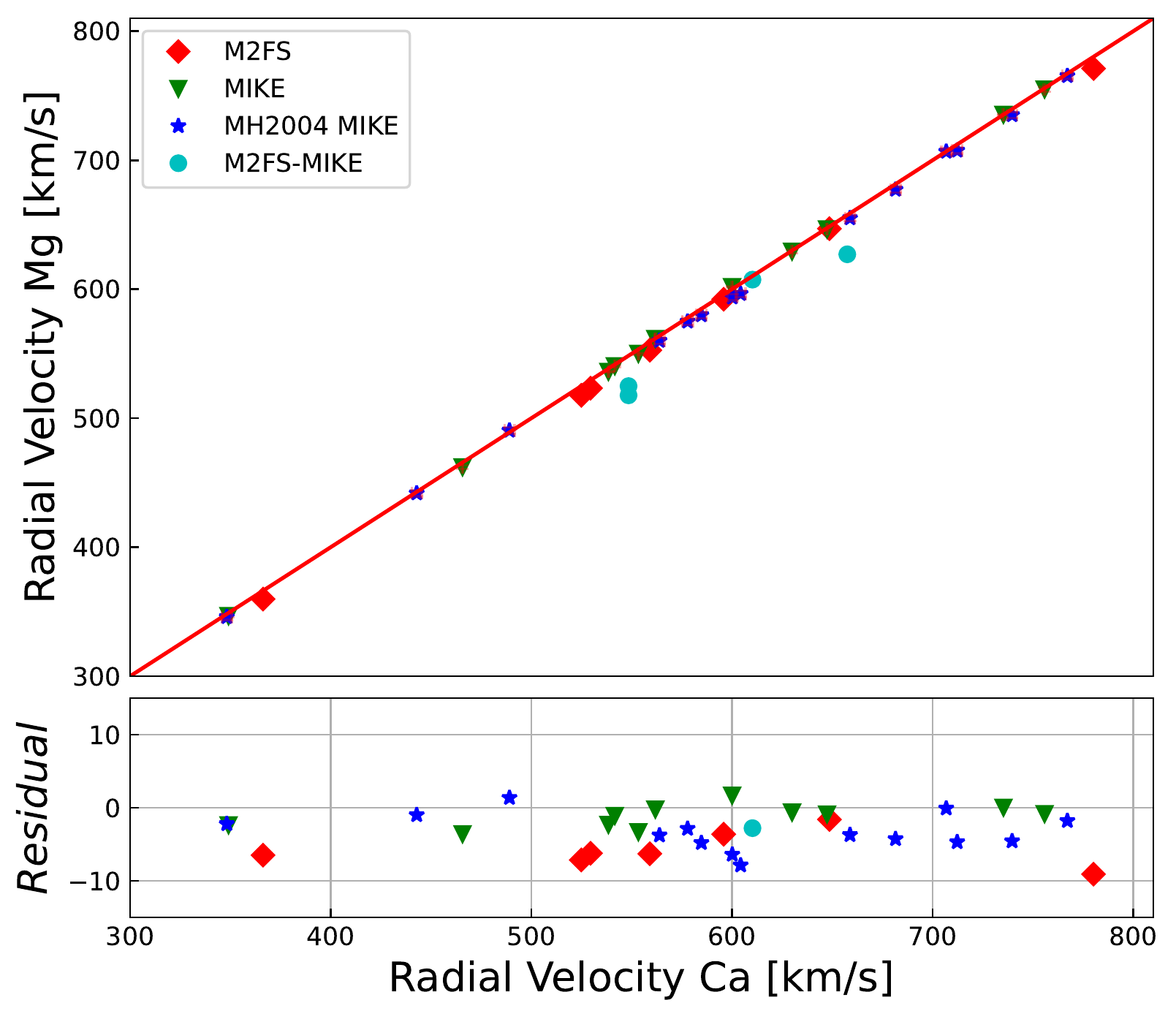}
\caption{Comparison of 37 radial velocities for luminous GCs observed both in the Calcium \& Magnesium Triplet. The red diamond and green inverse pyramids indicate GCs observed in both orders with M2FS and MIKE respectively.  The MH2004 symbols indicate our rereduction of the MIKE data from \citet{martini2004}, while the Cyan symbols indicate cross-instrument comparisons detailed in the text. 
\label{fig:internal_comp_RV}}
\end{figure}

Many targets were observed with multiple setups (M2FS Magnesium and Calcium Triplet), or both with M2FS and MIKE. This allows us to do a comparison of their radial velocity and velocity dispersion to test any possible instrumental bias or differences between the Calcium and Magnesium Triplet measurements. The radial velocity comparison for 23 luminous GCs observed in both the Magnesium vs Calcium Triplet is shown in Figure~\ref{fig:internal_comp_RV}. We also included the 14 luminous GCs from \citet{martini2004} (see \S\ref{subsec:spectral_fitting} for more details). For the Calcium triplet radial velocities we found a systematic offset toward higher values relative to Magnesium Triplet radial velocities. This systematic offset was $5.91$~km/s and $2.41$~km/s for objects observed with M2FS and MIKE respectively; because we have no way of knowing which velocity is correct, we add these offsets as errors in quadrature with the Monte Carlo radial velocity errors for the two instruments.

In comparing repeat measurements, we found that our Monte Carlo errors for our velocity dispersion (See \S\ref{subsec:spectral_fitting}) were significantly underestimated.  We therefore rescaled our errors based on these repeated measurements by fitting a Gaussian distribution to the difference of our repeated measurements normalized by their errors ($(\sigma_{Mg}-\sigma_{Ca})/\sqrt{error_{\sigma_{Mg}}^{2}+error_{\sigma_{Ca}}^{2}}$).  This gives us a scaling factor of 3.26 that we apply to all our velocity dispersion errors such that the repeated measurement differences normalized by the errors have a standard deviation of one. 

\begin{figure}
\plotone{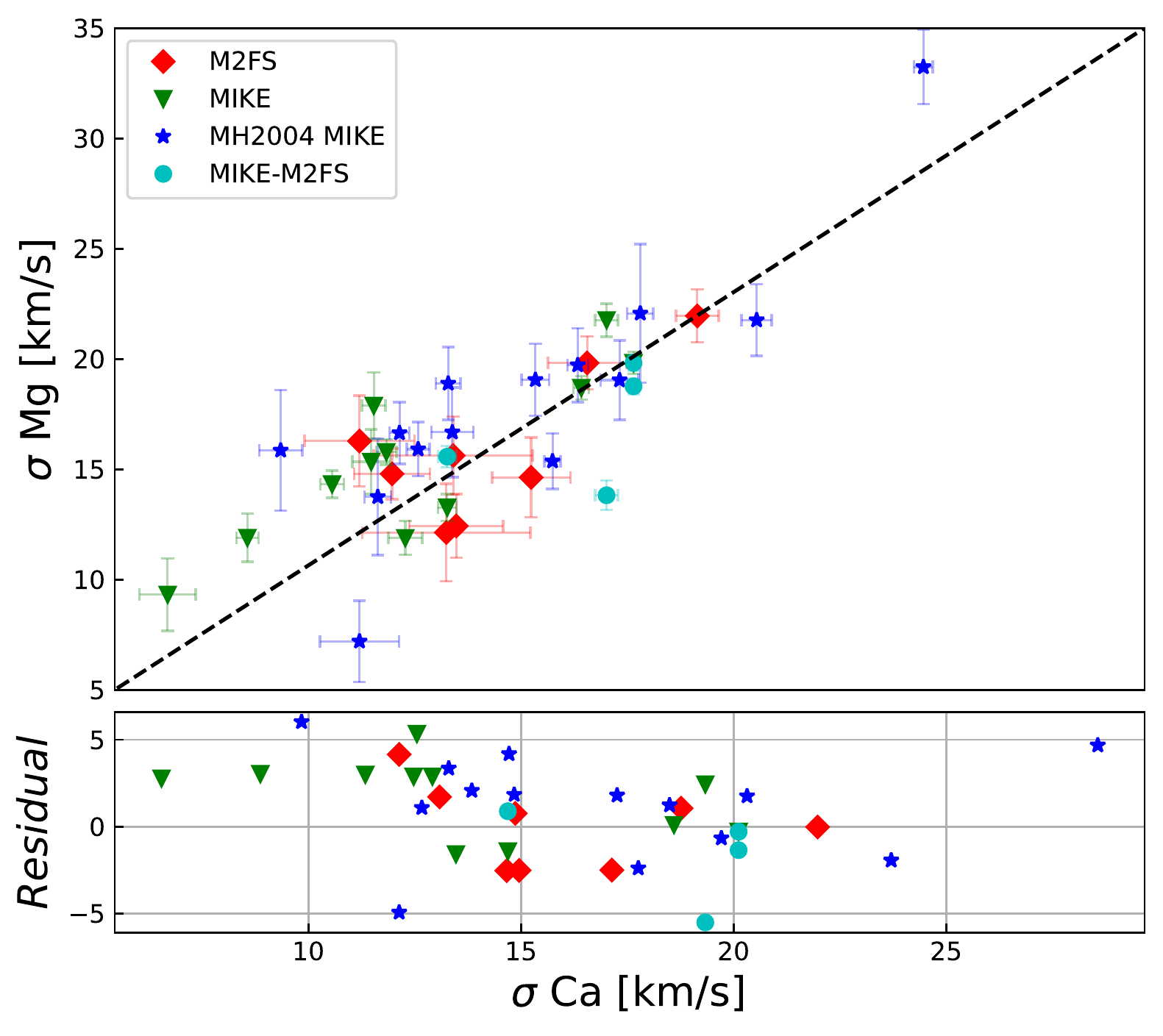}
\caption{ Comparison of 37 velocity dispersion measurements for luminous GCs observed both in the Calcium \& Magnesium Triplet. Symbols are as described in Figure~\ref{fig:internal_comp_RV}.
\label{fig:internal_comp_vdisp}}
\end{figure}

Three luminous GCs were observed with both M2FS and MIKE: \textit{H12\_78}, \textit{KV19-271}, and \textit{pff\_gc-098}. These are shown in cyan in Figure~\ref{fig:internal_comp_RV} and Figure~\ref{fig:internal_comp_vdisp} -- the x-axis position shows their MIKE calcium Triplet measurements. On the y-axis, for \textit{KV19-271}, the Magnesium and Calcium Triplet M2FS observations are plotted separately, while for \textit{H12\_78} we use the M2FS Calcium Triplet value.

Figure~\ref{fig:internal_comp_vdisp} shows the velocity dispersion comparison for the same 37 luminous GCs from Figure~\ref{fig:internal_comp_RV} observed both in the Magnesium and Calcium Triplet. We find a good agreement between the Magnesium and Calcium Triplet velocity dispersions, with a scatter of $\sim \: 2$~km/s.

\subsection{Metallicities}
\label{subsec:metallicities}
Using the weights of the templates used in the best fit from pPXF, we obtained a luminosity-weighted estimate of the metallicity of our luminous GC sample. The use of individual star templates (rather than Simple Stellar Population templates) and the known Age-Metallicity degeneracy for stars may lead to systematic uncertainties of the metallicity in some cases. For luminous GCs observed with MIKE where derived metallicities are available in the Calcium Triplet and also in the Magnesium Triplet (centered at $5127-5277$~\AA), we present the error-weighted average metallicity between the two measurements.

\begin{figure*}
\gridline{\fig{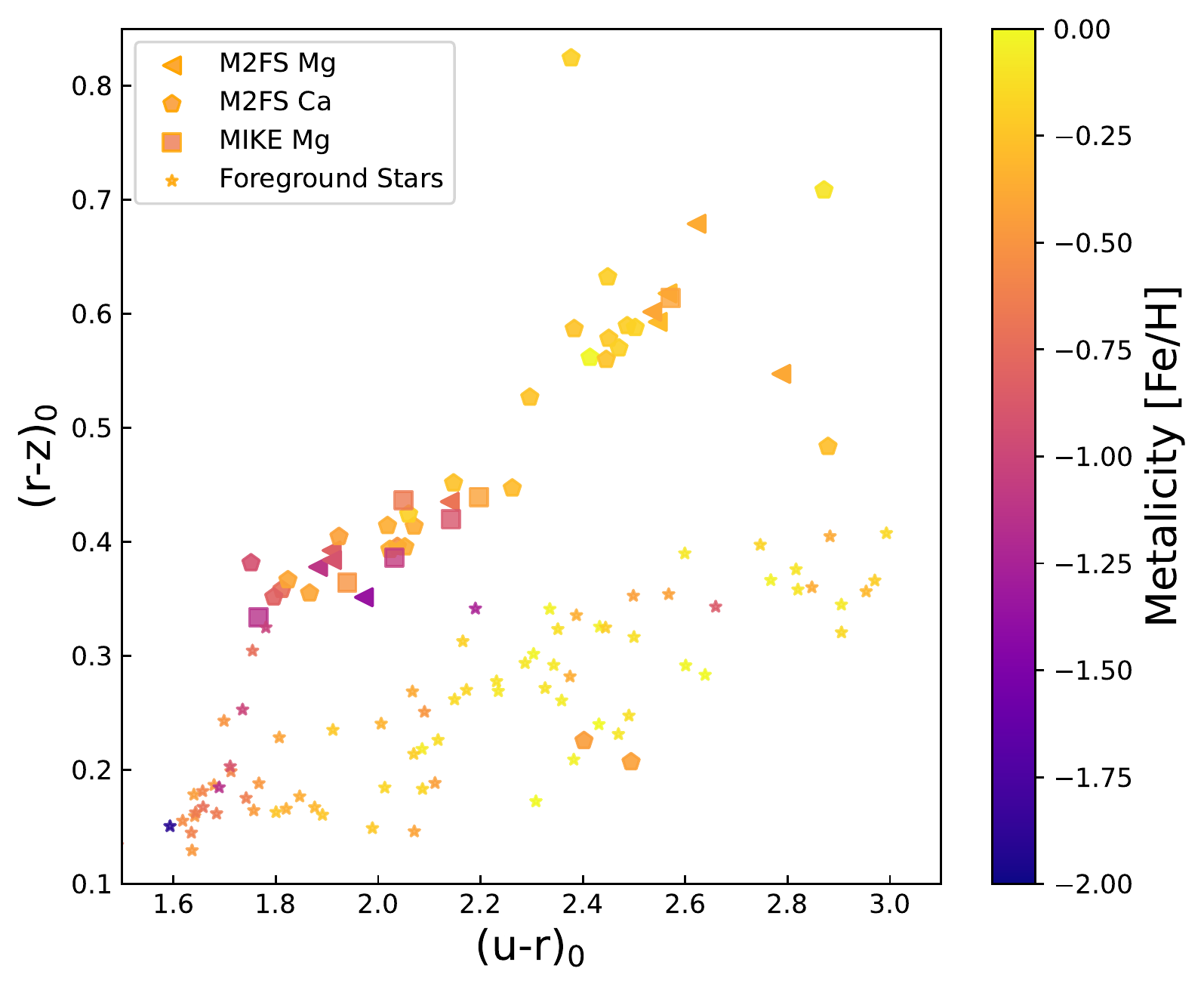}{0.5\textwidth}{}
        \fig{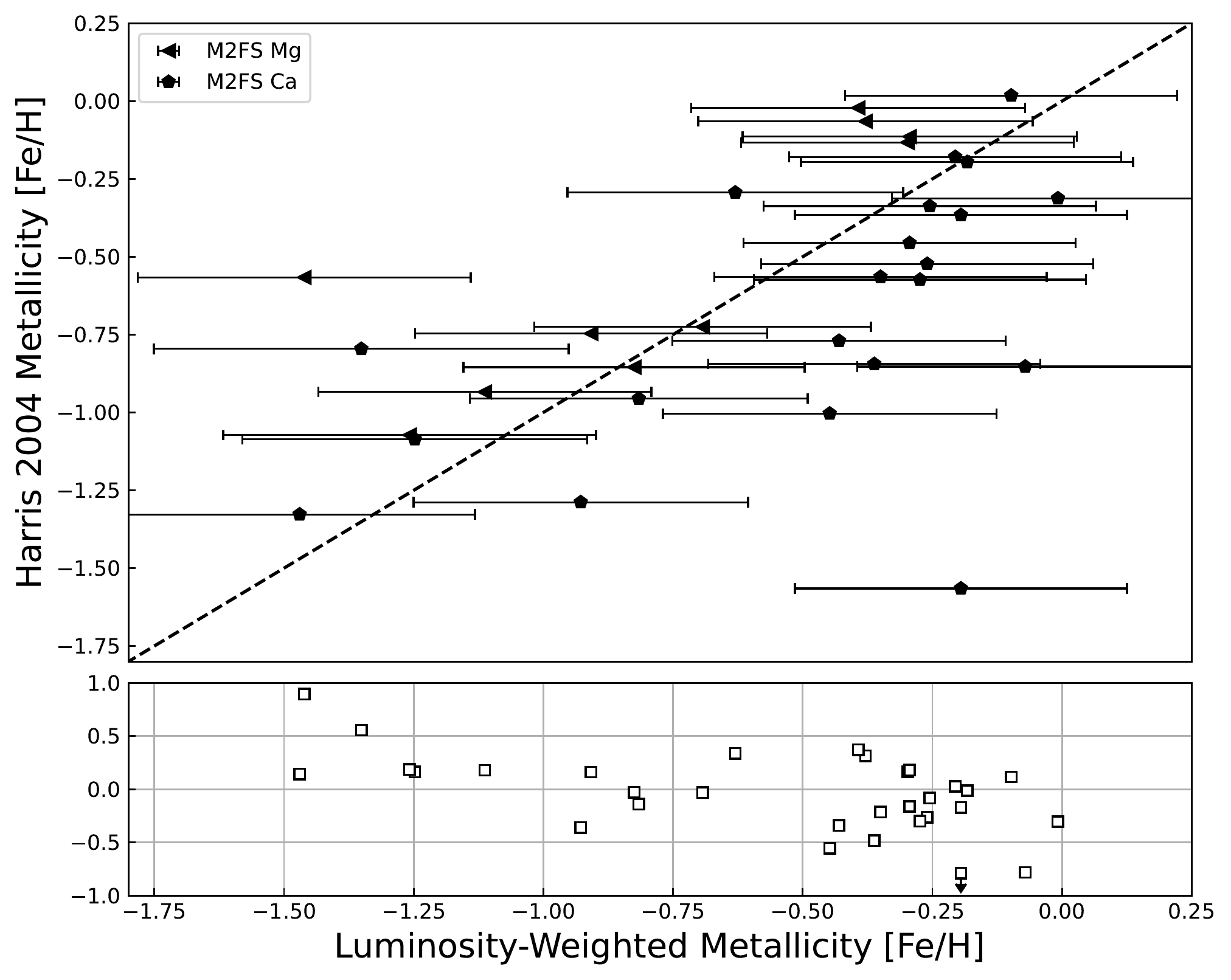}{0.5\textwidth}{}}
\caption{\textit{Left:}\textit{r-z} vs \textit{u-r} color plot for luminous GCs and Milky Way foreground stars colored by luminosity-weighted metallicities estimate. The triangles, pentagons and squares represents targets observed with M2FS Magnesium Triplet, M2FS Calcium Triplet and MIKE respectively .\textit{Right:} Comparison of our flux weighted metallicities (x-axis) to derived metallicites (y-axis) for 31 luminous GCs using color information from the Washington photometric system from \citet{Harris2004}. Symbols are as described above. The dashed black line shows the one-to-one comparison.
\label{fig:metallicities}}
\end{figure*}

To test the robustness of our metallicity estimates, we examine how they correlate with the measured colors from the NOAO photometry (\S\ref{subsec:sample_selection}), as well as metallicities for 31 of our luminous GCs inferred from Washington $C-T_{1}$ photometry from \citet{Harris2004}.  All 31 of these are M2FS objects.  These comparisons are shown in Fig.~\ref{fig:metallicities}. In the left panel we see the luminous GCs fall along the expected sequence with lower metallicity objects being bluer, while higher metallicity are redder.   In the right panel, the \citet{Harris2004} Metallicity estimates are well correlated with our estimated metallicities -- they have a Spearman coefficient of $r_{spearman}=0.56$ suggesting the correlation has a $>99\%$ significance, and a scatter of 0.32 dex. We have added this scatter in quadrature to the Monte Carlo errors in metallicity; these are the error bars shown in the right panel and listed in Table~\ref{tab:all_GC_data}.

{\em Outliers -- } in the right panel, we see two clusters lying at (u-r)$_0 \sim 2.4$ that fall far below the globular cluster sequence and have colors similar to foreground stars.   Both objects (\textit{K-029} \& \textit{HGHH-19}) have radial velocities consistent with being Cen~A GCs ($638.8\pm5.9$ km/s and $621.7\pm5.9$ km/s respectively), and no significant {\em Gaia} EDR3 proper motions ($\lesssim0.8\:mas/year$).  On the right panel, the largest outliers on the bottom right and top left (\textit{T17-2078} and \textit{HH-10}) are both located near the center of Cen~A, and both show close companions which contaminate their photometry or spectroscopic measurement and compromise the derived metallicity. Based on these outliers, and the large scatter found in the right panel of Figure~\ref{fig:metallicities} we suggest caution in any future use of individual GC metallicities published here.

\section{Radii and Luminosity Determination}\label{sec:radii_lumin}

In this section we discuss how we derived structural parameters and determine $V$-band luminosities for our sample of $57$ luminous GCs with visually verified velocity dispersions.
Structural parameters, in particular half-mass radii, are needed to derive virial masses.  We then use the virial masses combined with $V$-band luminosities to estimate their mass-to-light ratios, which is the primary scientific goal of this work.

\subsection{Radii Determination}\label{subsec:radii}
Generally, half-mass radii are determined from the modeling of the light profile of imaging data, such as King models \citep{King62,King66} hereafter King62 and King66 respectively. 
At the distance of Cen~A the typical globular cluster half-light radius of 3~pc corresponds to just 0$\farcs$16,  thus high resolution imaging is needed to estimate their sizes. Unfortunately, there is  only archival high resolution HST imaging data for twelve of our luminous GC candidates.  We also have ground-based PISCeS imaging at the $6.5$~m telescope Magellan Clay \citep{Crnojevic2016} -- while these data were taken in good seeing conditions ($\sim 0.\arcsec65$), they do not resolve the GC sizes, and modeling of the sizes from PSF convolved-fits result in large errors in the inferred sizes.  

For that reason, to determine the physical sizes of our GCs candidates we revisited the relation found in \citet{KV2020} between HST-based luminous GC sizes and the BP-RP {\em Gaia} excess factor (hereafter $BR_{\rm excess}$). The $BR_{\rm excess}$ is the ratio of the combined fluxes of the Blue-Pass (BP, $3300-6800$~\AA) and Red-Pass (RP,$6400-10500$~\AA) calculated in a window of $3.5\arcsec \times 2.1$\arcsec~ with the G band flux determined from PSF fitting ({\em Gaia} EDR3 has an effective angular resolution of $\sim0.\arcsec4$; \citealt{Fabricius2021}). The $BR_{\rm excess}$ is sensitive to extended objects as the BP and RP are calculated over a larger area than the PSF photometry of the G band. The BR$_{\rm excess}$ is $\sim$1 for point sources and larger for extended sources. We decided to use the newly available {\em Gaia} EDR3 catalog, instead of {\em Gaia} DR2 used in \citet{KV2020}. {\em Gaia} EDR3 is more complete and thus has more sources with existing HST measurements for comparison.  Additionally, the BR$_{\rm excess}$ has also been recalibrated in {\em Gaia} EDR3 \citep{GaiaEDR3_phot} relative to the {\em Gaia} DR2 values presented in \citet{KV2020}; we also use half-mass radii here in place of half-light radii used in \citet{KV2020} to provide the inputs needed for virial masses.  

\begin{figure*}
\gridline{\fig{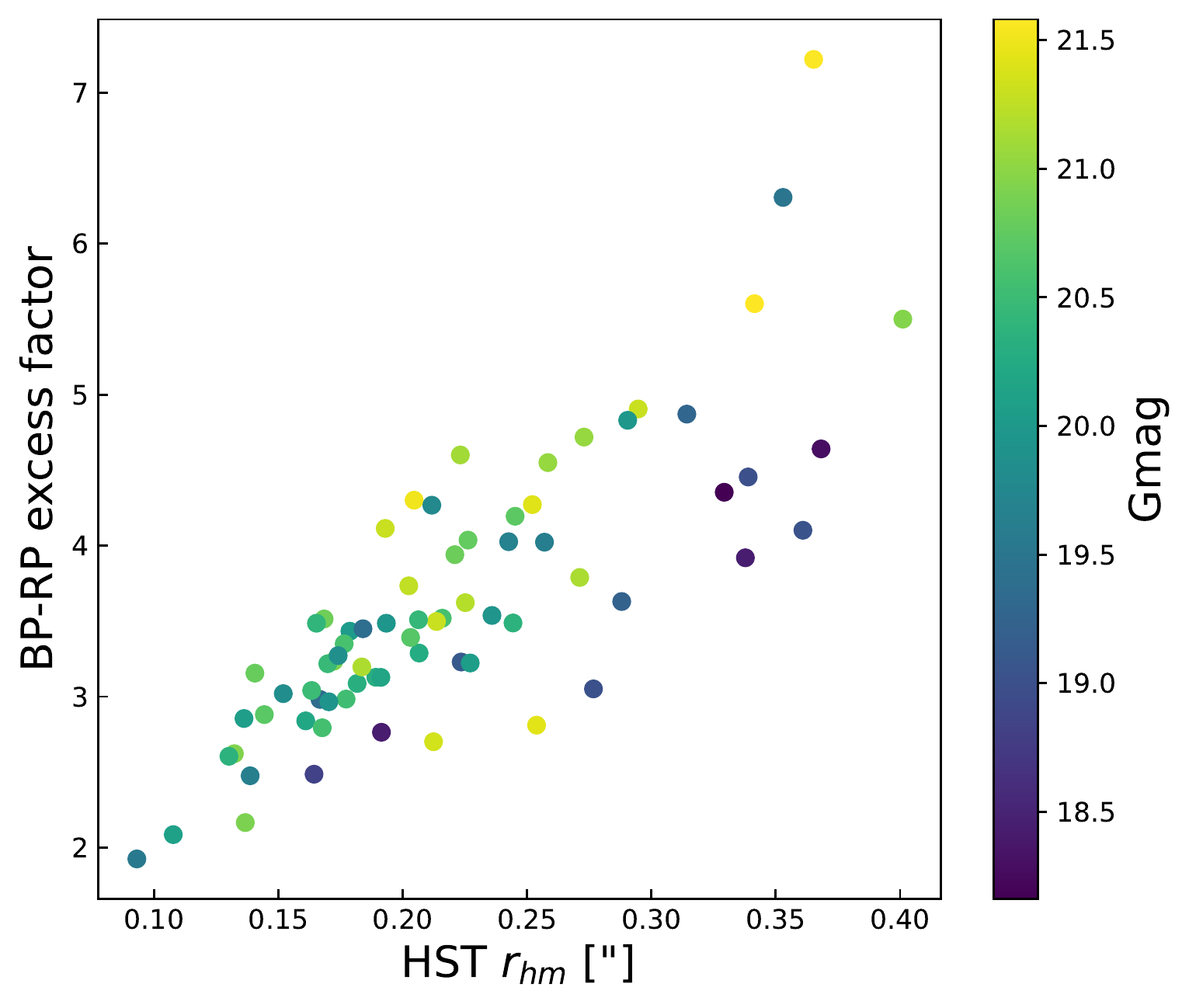}{0.43\textwidth}{}
        \fig{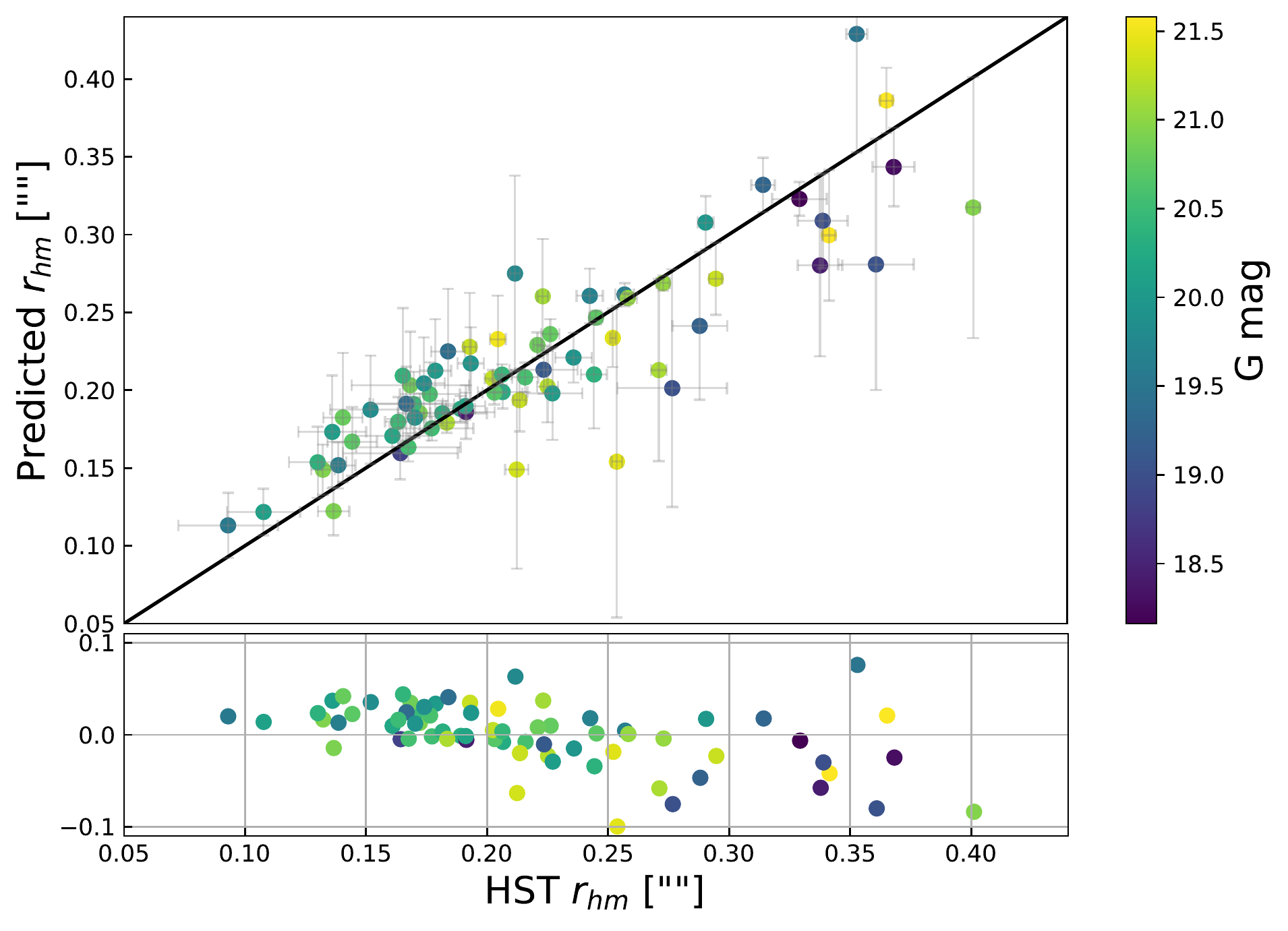}{0.5\textwidth}{}}
\caption{\textit{Left:} A correlation is seen between the {\em Gaia} EDR3 BR$_{\rm excess}$ and the half-mass radii for Cen~A GCs with available literature sizes from \citet{McLaughlin2008}. \textit{Right} Predicted half-mass radii for the same luminous GCs shown in the left using the fit to the correlation in the left panel given in Equation~\ref{eq:1}. This relation has a scatter of $14$~\%.\label{fig:predicted}}
\end{figure*}

To determine the relation between half-mass radius and $BR_{\rm excess}$, we compiled a list of 71 half-light radii ($r_{h}$) and concentrations ($c=r_{t}/r_{c}$; with $r_{t}$ and $r_{c}$ being the tidal and core radii) of luminous GCs in Cen~A derived from King66 model fits to HST images from  \citet{McLaughlin2008}. 
To convert their half-light radii to half-mass radii, we integrated King66 models for a range of concentrations to obtain the following relation:
\begin{equation}
    \label{eq:rh_to_rhm}
    \begin{aligned}
& r_{hm}= r_{h}(1.33592 - 0.05677\log(c) + 0.03942\log(c)^{2}\\
&    +0.01146\log(c)^{3} -0.00622\log(c)^{4})
    \end{aligned}
\end{equation}
The left panel in Figure \ref{fig:predicted} shows the resulting half-mass radii vs the measured {\em Gaia} BR$_{\rm excess}$ colored by {\em Gaia} G magnitude. These two quantities show a strong positive correlation with a Spearman coefficient r$_{spearman}=0.81$ suggesting the correlation has a $>99.9$~\% significance.

We find a better recovery of the half-mass radii when including also a G band magnitude dependence in the relation between half-mass radii and $BR_{\rm excess}$.The resulting relation is shown in equation \ref{eq:1}. 

\begin{equation}
    \label{eq:1}
    \begin{aligned}
& r_{hm}(")=(0.08385-0.0086(G-18))(BR_{excess}-2.5)\\
& -0.00785(G-18)+0.16135
    \end{aligned}
\end{equation}

Right panel in Figure \ref{fig:predicted} shows the predicted half-mass radii for the sample of literature GCs. We find a fractional scatter of $14\%$, which is superior to the quality of ground based sizes as well as the relation found in \citet{KV2020}.

In this work we use half-mass radii from King66 fits to HST data when available \citep{McLaughlin2008,Georgiev2009}, and sizes from {\em Gaia} EDR3 from equation \ref{eq:1} otherwise to calculate the mass-to-light ratios for our 57 luminous GCs.  In addition three luminous GCs \textit{Fluffy}, \textit{vhh81-5}, and \textit{VHH81-01} as well as galaxy nuclei \textit{ESO269-06} were not in {\em Gaia} EDR3. For both \textit{VHH81-5}
and \textit{ESO269-06} HST ACS F606W imaging was available, and we fit King62 profiles (which are analytical) to determine half-mass radii and concentrations using a custom IDL routine that includes PSF convolution \citep[as in][]{Seth2010m32}.  For \textit{Fluffy}, due to its large size and resolution into individual stars, we created a 1-D surface brightness profile from the HST WFC3 F606W, and fit this to a King62 model; this work will
be presented in more detail in Crnojevic et al. {\em in prep}. We demonstrate in \S\ref{subsec:BHMass} that the differences between King62 and King66 profiles are negligible for the purposes of determining dynamical masses.  Finally, for  \textit{VHH81-01} no HST data was available, however, due to its large size, it was easily resolved in our ground-based PISCeS imaging.  Using the $r$ band image, we fit the cluster with PSF convolution to a King66 model using {\tt ishape} \citep{Larsen1999}; the seeing of the $r$ band data was 0$\farcs$5, significantly smaller than the $\sim$1.3'' half-light radius (See more details in ~\S\ref{subsec:posa_20}).

\subsection{$V$-band Luminosities}
\label{subsec:luminosities}
To make our mass-to-light ratios comparable to previous work, we derived $V$-band mass-to-light ratios. This requires that we transform our extinction corrected NOAO Source Catalog DR2 aperture photometry (see \S\ref{subsec:sample_selection}) in $u$, $g$, $r$, \& $z$ into $V$ band using  the photometric transformation: ${\rm V} = {\rm g} - 0.59({\rm g-r}) - 0.01$ 
from \citet{Jester2005}. We compared the derived $V$-band magnitudes with previous $V$-band magnitudes for 185 GCs in Cen~A from \citet{Peng2004} and \citet{Rejkuba2001}, and saw a slight bias of $0.087$ mag towards brighter magnitudes in the NOAO DR2 photometry; this offset was constant with magnitude. Comparisons between \citet{Rejkuba2001} and \citet{Peng2004} photometry suggests they are consistent, thus the offset appears to be related to a photometric calibration or aperture correction issue with the NOAO DR2 catalog.  We note that available PISCeS photometry is saturated for many of our targets.  We therefore use the derived $V$-band magnitudes derived from NOAO DR2 and subtract $0.087$ mag to correct for this bias. For 10 objects we used HST imaging to obtain $V$-band magnitudes, these are detailed in Appendix~\ref{app:magnitudes}. 

For all 57 clusters, we calculated the $V$-band luminosities using $L_{V}=10^{-0.4*(V-(m-M)_{V}-A_{V}-M_{V,\odot})}$; we assumed M$_{V,\odot} = 4.81$,  $(m-M)_{V}=27.91$ (i.e. distance of 3.8 Mpc) from \citet{Harris2010}, and $A_{V}$ was calculated using \citet{Schlafly2011} for each object individually. The absolute $V$-band magnitudes $M_{V}$ and E(B-V) values used are listed in Table~\ref{tab:good_GC_data}.

\section{Mass-To-Light Ratios}\label{sec:M/L}

A primary goal of this work is to find potential stripped nuclei around Cen~A that have elevated mass-to-light ratios that indicate the presence of a massive black hole \citep{Mieske2013,Seth2014m60,Karina2019}.  
We use the velocity dispersion measurements combined with the half-light mass radii estimates from section \ref{subsec:radii} to calculate each luminous GCs virial mass using equation 1 from \citet{Strader2009},

\begin{equation}
    \label{eq:2}
    \begin{aligned}
& M_{vir} = \frac{7.5 \sigma^{2}_{\infty} r_{hm}}{G}
    \end{aligned}
\end{equation}

\noindent where $\sigma_{\infty}$ is the global velocity dispersion and $r_{\rm hm}$ is the half-mass radius (see \S\ref{sec:radii_lumin}), and $G$ is the gravitational constant.  
At the distance of Cen~A, luminous GC candidates are partially resolved from ground-based observations, resulting in the derived velocity dispersion being different from the global velocity dispersion ($\sigma_{\infty}$). Due to the declining dispersion with radius in globular clusters, $\sigma_{\infty}$ will always be lower than our measured velocity dispersion. We calculate this difference by integrating a King66 profile convolved with the seeing over the spectrograph aperture of our observations as previously done in \citet{Strader2009}. For clusters with {\em Gaia} estimated sizes, we assumed a concentration parameter $c = 30$, a typical value for Local Group globular clusters. The typical difference between the derived and global velocity  dispersion is about 2\% at $r_{\rm hm} = 2.7$~pc, 5 \% at $r_{\rm hm} = 4$~pc and 7\% for $r_{\rm hm} = 7$~pc, with 1~pc $\approx 0.05\arcsec$. The maximum correction for any cluster in our sample was 15\% for cluster \textit{VHH81-01} which has a half-mass radius of 1.78\arcsec. Both our derived dispersion and the calculated $\sigma_{\infty}$ are listed in Table~\ref{tab:good_GC_data}. 

\begin{figure}[h!]
\plotone{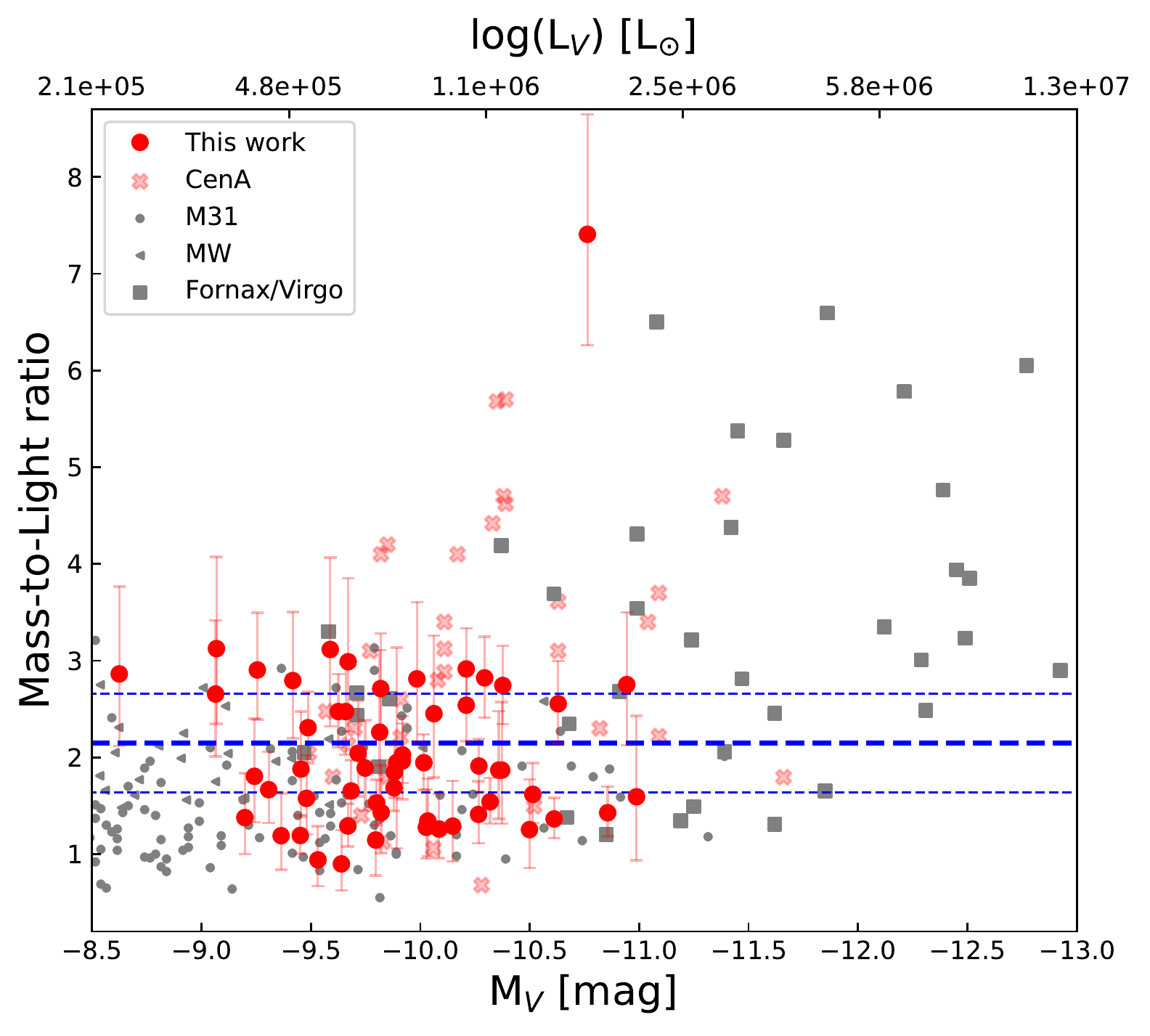}
\caption{ Luminosity of stripped nuclei and GCs compared to their M/L$_V$. Solid red circles are the 57 objects from this paper, while lighter red points are Cen~A object from the literature and gray points are from other galaxies (see text for details).  The horizontal dashed lines are the best-fit and scatter of the stellar mass-to-light ratios of long relaxation systems from \citet{Karina2019}.  
 \label{fig:Mass-to-light-ratios}}
\end{figure}

Figure \ref{fig:Mass-to-light-ratios} shows the $V$-band mass-to-light ratios (M/L$_V$) for our sample of 57 luminous GCs in Cen~A.  Also shown are GCs from M31 \citep{Strader2011}, UCDs from Virgo and Fornax compiled by \citet{Mieske2013}, as well as literature objects from Cen~A \citep{martini2004,Rejkuba2007}.  Our mass-to-light ratios were computed by dividing the derived virial masses by the calculated $V$-band luminosities. For the CenA's literature objects from \citet{martini2004} and \citet{Rejkuba2007} we use updated half-light radii from \citet{McLaughlin2008} to revise the $M/L_V$ estimates presented in the original papers. The half-light radii are then converted to half-mass radii (necessary to get $M/L_V$) using equation~\ref{eq:rh_to_rhm}. The list of clusters, half-mass radii, and their M/L$_V$ values are shown in Table~\ref{tab:Rejkuba}. As discussed in \S2, we do not include objects from \citet{Taylor2010} and \citet{Taylor2015} in Fig.~\ref{fig:Mass-to-light-ratios} due to their typically low S/N.  We also note that the half-light radii used in \citet{Taylor2015} are from unpublished work and are systematically larger than those presented by \citet{McLaughlin2008} -- this would bias these values to higher $M/L_V$ values than those presented in our work.  

\begin{figure}
\plotone{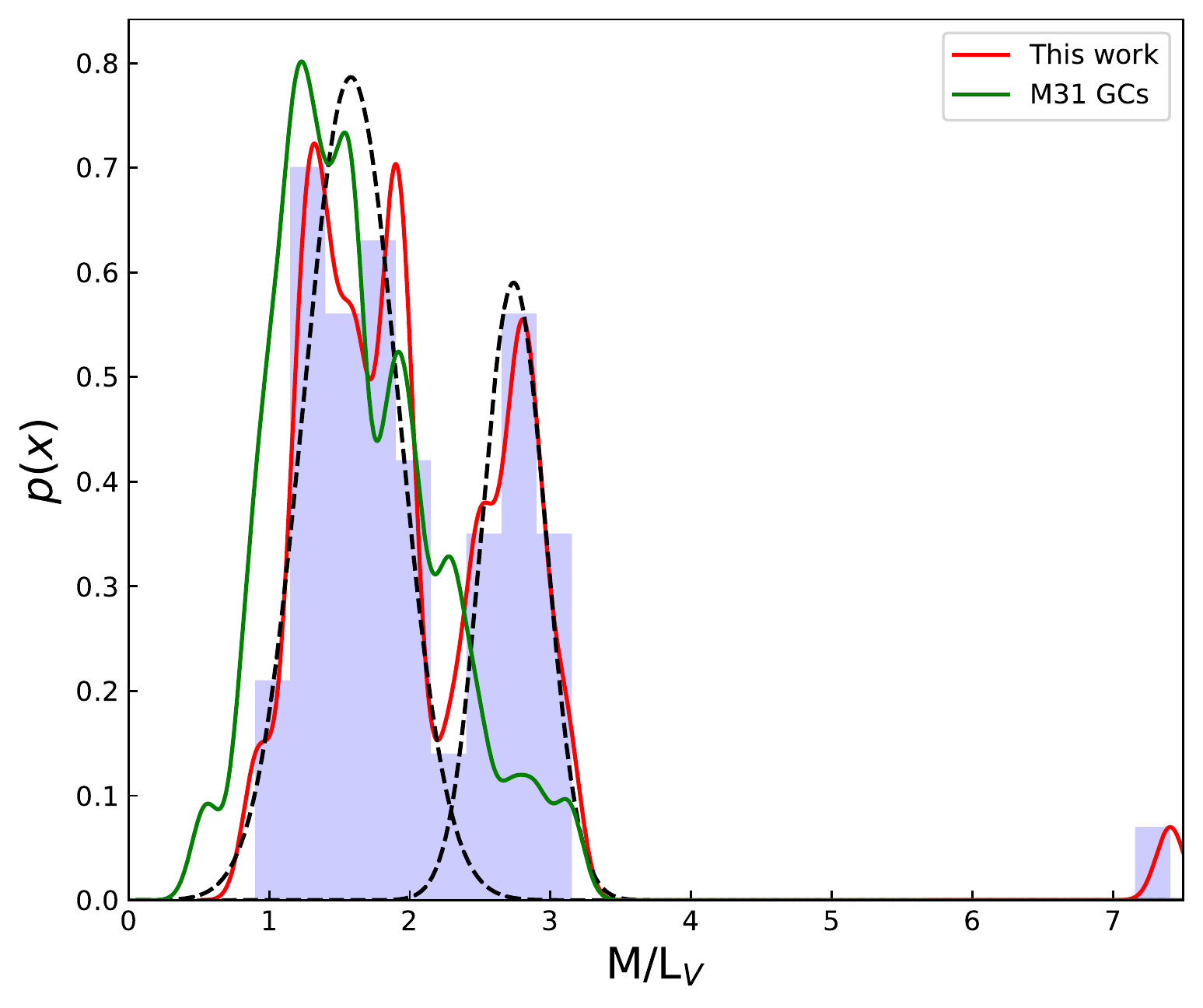}
\caption{Histogram of the distribution of the M/L$_V$ for the 57 luminous GC. A clear bimodality is seen in our data. A one-dimensional Kernel Density Estimation for our luminous GC is shown in red, compared with a similar estimate for M31 globular clusters (shown in green). The black dashed lines shows the two mixture models for the best Gaussian mixture model for our luminous GCs. We find a clear bimodality with ``normal'' GCs peaking at M/L$_V=1.51$, and luminous GCs with inflated M/L$_{V}$ peaking at M/L$_V=2.68$. The two populations are divided at M/L$_V=2.3$. The outlier in M/L$_V$ is our most massive luminous GC \textit{VH81-01} (More details \S\ref{subsec:posa_20}). \label{fig:Hist_ML}}
\end{figure}

To investigate the M/L$_V$ trends further, we plot their histogram in Fig.~\ref{fig:Hist_ML} (blue shaded region), revealing a bimodal distribution. We also show a kernel density estimate of the same data with a kernel width of 0.1 in red, while a similar kernel density estimate for the $45$ M31 clusters with M$_V < -9.5$ from \citep{Strader2011} are shown in green. We choose to only include the brightest M31 clusters in this comparison to make the sample comparable in brightness to our Cen~A clusters, and to reduce the impact of M/L evolution due to relaxation in the lower mass clusters \citep[e.g.][]{Kruijssen2008}. The bulk of clusters in both samples have $M/L_V$ between 1 and 2.  However, in Cen~A, the histogram as well as the comparison with M31 reveal a second population of clusters with an elevated $M/L_V \sim 3$. A Kolmogorov–Smirnov test gives a $>99.6 \%$ significance that our observed distribution of M/L$_V$ for luminous GCs in Cen~A is different from those of M31 GCs.  We also note that the high $M/L_V$ population in Cen~A is above the  mean value of $\sim 2.15$ inferred by \citet{Karina2019} from long relaxation M31 and Milky Way globular clusters and existing Virgo UCD measurements with black hole components removed.

To quantify the fraction of our luminous GCs with elevated M/L$_V$ we used a Gaussian mixture model using the {\tt sklearn} Python package\footnote{\url{https://scikit-learn.org/stable/modules/generated/sklearn.mixture.GaussianMixture.html}}. After excluding the highest M/L$_V$ cluster \textit{VHH81-01} (our most massive and largest luminous GC, more details \S\ref{subsec:posa_20}), we find that a Gaussian mixture model with 2 components best describes the observed distribution of M/L$_V$, with the lower M/L$_V$ population peaking at M/L$_V=1.51$ with standard deviation of $0.32$. The higher M/L$_V$ peak is at $2.68$ with a standard deviation of $0.21$.  The Gaussian mixture model suggests that 35.2\% of our sample of luminous GCs (20/57) have an inflated M/$L_V$.  This is the same as the number of objects that are above M/$L_V$ of 2.3, which represents the dividing line where the probability of being in the lower and upper M/$L_V$ population are equal. Relative to M31, we find $2.53\times$ more luminous GCs above $M/L_V$ of 2.3.

One expectation from stellar models that is not borne out in observations is that the M/L$_V$ should be metallicity dependent; this has also been found in previous work \citep{Strader2011, Bruzual2013}. We find no significant evidence of any trend of  M/$L_V$  with our derived metallicity (See \S\ref{subsec:metallicities}) or PISCeS $g-r$ color. We note that for two objects, \textit{KV19-295} and \textit{T17-1648}, our M/L$_V$ values may be incorrect due to bad $g-r$ colors used to estimate their $V$-band luminosity (see Section~\ref{app:magnitudes}); their M/L$_V$ values are 1.6 and 1.1, and thus both belong to the lower $M/L$ population.

\subsection{Nuclear Star Clusters}  
Our observations include four nuclear star clusters of present day galaxies.  Of these, we were able to estimate mass-to-light ratios in three: \textit{KK197-NSC}, \textit{ESO269-06} and \textit{DW3} from \citet{Crnojevic2016}. All three galaxies have total $V$-band luminosities $<$4$\times$10$^7$~L$_\odot$ \citep{Crnojevic2019} and thus stellar masses $\lesssim$10$^8$~M$_\odot$.  The nuclear star clusters have M/L$_V$ of $1.13$, $1.69$ and $1.15$ respectively, and thus none are consistent with being elevated and having high mass fraction black holes. Their virial masses range from 6$\times10^{5}$M$_{\odot}$ in \textit{DW3} to 1.2$\times10^{6}$M$_{\odot}$ in \textit{ESO269-06}.
We note that \textit{KK197-NSC} has a dynamical mass-to-light ratios estimate from \citet{Fahrion2020}. We find a $\sim3.5\times$ lower M/$L_V$ than their published value of M/$L_V=5.65\pm0.5$. This disagreement is due solely to their derived global velocity dispersion of $24.8\pm0.8$ km/s from VLT UVES data, which is a factor of $\sim2\times$ higher than our global velocity dispersion of 12.5$\pm$0.4 km/s. For comparison we fitted the MIKE Calcium Triplet spectrum of \textit{KK197-NSC} with a fixed velocity dispersion of $24.8$~km/s from \citet{Fahrion2020}. The fit has a reduced $\chi^{2}$ of 1.48, this is $0.06$ higher than our best fit reduced $\chi^{2}$ of 1.42, a difference of more than 3$\sigma$. Additionally, the velocity dispersion measurement for the (lower S/N) MIKE Magnesium Triplet spectra of 15.3$\pm$4.3 km/s is consistent with our best fit MIKE Calcium Triplet velocity dispersion within the errors. We believe that the difference in velocity dispersion arises from the difference in extraction apertures.  We use MIKE data with a $1$\arcsec~slit diameter and an optimal extraction with FWHM of $0.81$\arcsec. The extraction aperture in \citet{Fahrion2020} was significantly larger, $1.\arcsec2\times4.\arcsec0$, and thus may be dominated by the kinematics of the galaxy. Thus overall, we find that none of the nuclei have inflated mass-to-light ratios.  Either these nuclei have no central black holes, or their black holes are lower mass than we can detect.  As we show below, we can only detect black holes with $\gtrsim $10$^4$~M$_\odot$ (or $\gtrsim$4\% of the cluster mass).  

\subsection{BH Mass Estimates for Luminous GCs}
\label{subsec:BHMass}
The population of high M/L$_V$ objects could be due to the presence of massive black holes.  This idea was suggested by \citet{Mieske2013}, and the presence of black holes indicated by enhanced M/L$_V$ values was verified in five high mass UCDs in Virgo \citep{Seth2014m60,Afanasiev2018,Ahn2017,Ahn2018} as well as in the nuclei of galaxies \citep{Pechetti17,Krajnovic2018}.  The finding of high mass fraction black holes in UCDs/luminous GCs is a strong indication of them being stripped galaxy nuclei.  In this section we provide an estimate for the hypothetical central black hole (BH) mass for the 20 luminous GCs with elevated mass-to-light ratios (M/L$_{V} > 2.3$). The basic assumption we make here is that the inflated M/L$_{V}$ values are due to the presence of a black hole.  We create dynamical models to derive the black hole mass required in each object assuming the true stellar M/L$_{V}$ is given by the peak of the lower Gaussian mixture model component of 1.51 (see Section~\ref{sec:M/L}).  This then provides a rough estimate of the masses of the possible black holes in these systems. 

We first created a synthetic one-dimensional King62 light profile for each object (due to the availability of an analytic expression for the King62 models).  For objects with HST measurements we used their corresponding concentration ($c$) and half-light radii, while for those where our half-light radii are from {\em Gaia}, we assume a concentration parameter $c=30$.  We then fit a Multi Gaussian Expansion (MGE) model \citep{MultiGaussian_Cappellari} to the one-dimensional King62 light profiles. Next, we use the MGE models as an ingredient in Jeans dynamical models \citep[as implemented in the spherical version of the Jeans Anisotropic Modeling software JAM;][]{JAM_cappellari}. These models use the MGE light profiles of the clusters to predict the velocity dispersion within the observed aperture as a function of two free parameters: the stellar mass-to-light ratio and black hole mass.  For this modeling, we assume the orbits of stars in the cluster are isotropic.  Comparison of the predicted dispersion to the observed dispersion enables constraints on the mass-to-light ratio and BH mass.  We note that without a black hole, the derived M/L$_V$ for the clusters agrees extremely well with the virial mass measurements presented above (well within the 1$\sigma$ errors), suggesting the differences created by translating the effective radii and concentrations of King66 to King62 models are negligible.  
Next, we fixed the stellar mass-to-light ratio to $1.51$, the mean M/L$_V$ for clusters with lower M/L$_V$ (see~\S\ref{sec:M/L}). Using this stellar M/L$_V$, we run a set of JAM models with a grid of increasing BH masses to simulate the effect of a BH on the integrated velocity dispersion. The BH mass that gives the closest velocity dispersion to the observed value is picked as the best BH mass. We obtained an upper and lower limit for the predicted BH mass using the errors in the M/L$_V$ of each individual luminous GC. 
  
\begin{figure*}
\gridline{\fig{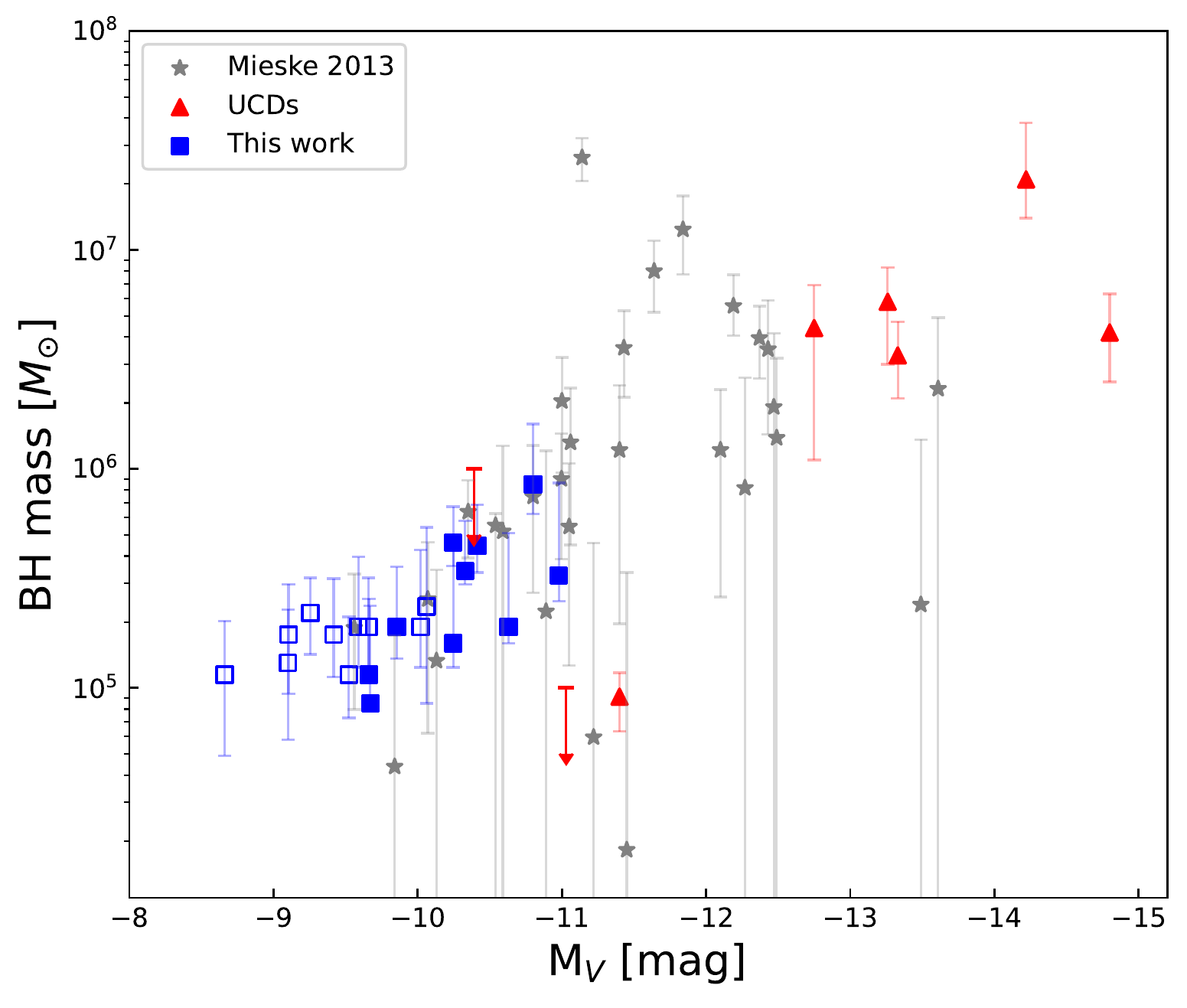}{0.5\textwidth}{}
        \fig{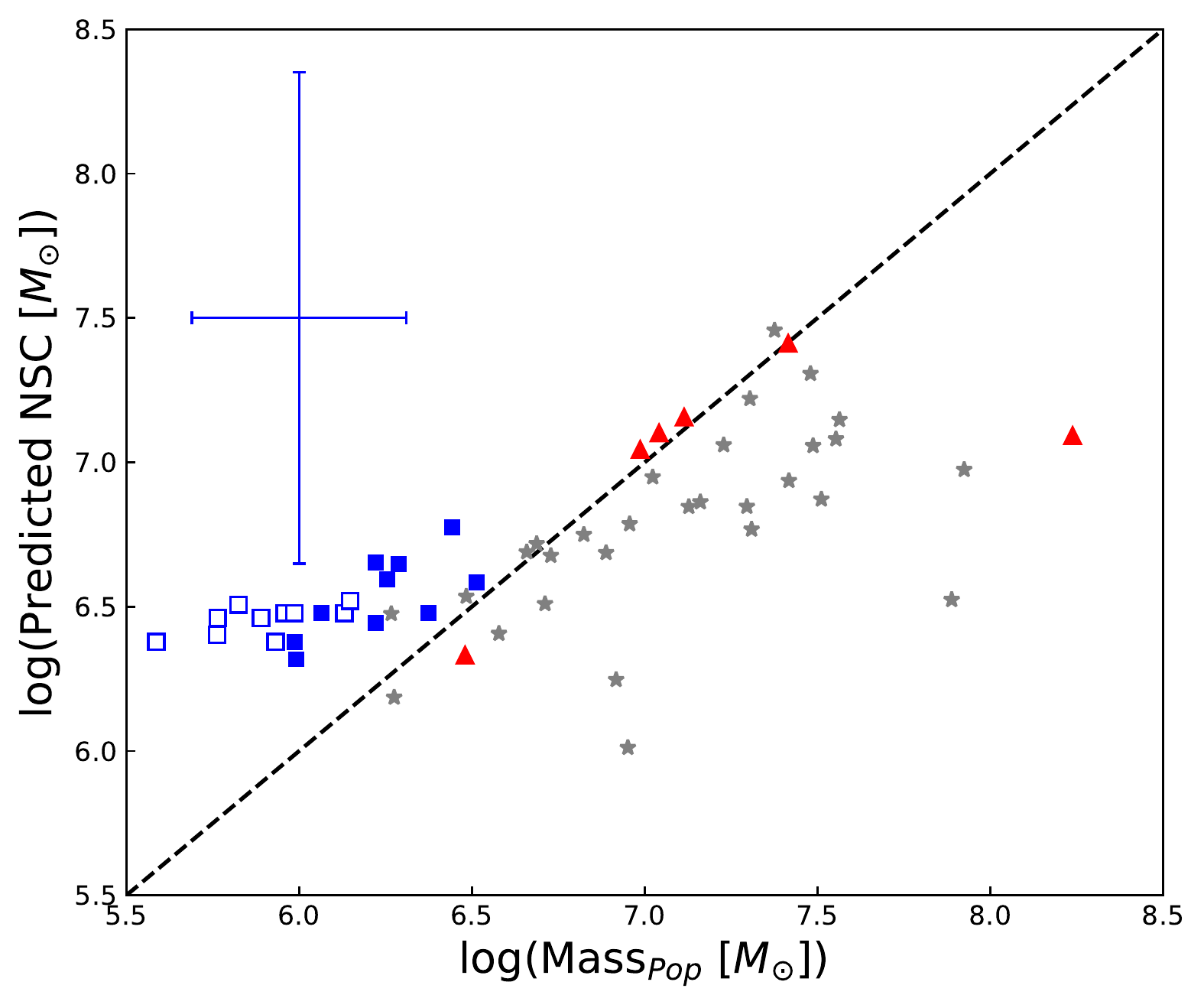}{0.505\textwidth}{}}
\caption{\textit{Left:} Predicted black hole masses for 20 luminous GCs in Cen~A with inflated mass-to-light ratios vs.~their absolute $V$-band magnitudes (blue squares). Solid square show luminous GCs with M/L$_V$ enhancements $>3 \sigma$, while open squares have less significant M/L$_V$ enhancements.  Gray stars are black hole mass predictions for stripped nuclei from \citet{Mieske2013}. The five known black hole masses in UCDs from \citet{Seth2014m60,Ahn2017,Ahn2018,Afanasiev2018} are shown in red triangles together with two upper limit for two UCDs from \citet{Karina2018} and the recent BH mass estimate from the most massive M31 globular cluster (B023-G078; M$_{V}=-11.4$) by \citet{Pechetti2021}.  \textit{Right:} The X-axis shows the mass for 20 luminous GCs based on their $V$-band luminosity (L$_{V}$) from Table~\ref{tab:good_GC_data} assuming a M/L$_{V}=1.51$ (Mass$_{pop}=$~L$_{V}\times1.51$). The Y-axis shows the predicted NSC mass based on the inferred black hole in the left panel using the empirical relation between black hole and total galaxy stellar mass \citep[][; eq. 4]{Reines2015} and total galaxy stellar mass and NSC mass \citep[][; eq. 5]{Neumayer2020}. The blue error bar indicates the uncertainty in both quantities: the vertical error bar is dominated by the scatter in the scaling relations, and the horizontal error bar is determined based on the width of the lower M/L$_{V}$ component in Figure \ref{fig:Hist_ML}.  Objects close to the one-to-one line (black dashed) are consistent with being stripped galaxy nuclei.
\label{fig:BHmass}}
\end{figure*}

The predicted BH masses vs. their $V$ magnitude are shown in the left panel of Figure~\ref{fig:BHmass} and their BH masses are listed in Table~\ref{tab:good_GC_data}. We distinguish luminous GCs with significant BH mass predictions (BH mass $3\times$ above their lower error) with solid blue squares from less significant BH mass predictions (open blue squares). We also plot BHs determined in a similar way from \citet{Mieske2013}. Our luminosity-BH relationship follows a similar trend to that found in \citet{Mieske2013} extending the predicted BH mass for luminous GCs down
to M$_{V}$ fainter than -9.5 (M$_{V} > -9.5$).

We find that BH masses of $4-18$~\% of the luminous GC virial mass (as calculated in Section~\ref{sec:M/L}) explain the inflated M/$L_V$ that we observe, with a mean BH mass of $3.3\times10^{5}$~M$_{\odot}$ and a maximum BH mass of $8.61^{+10.3}_{-0.12}\times10^{5}$~M$_{\odot}$ for the luminous GC \textit{VHH81-01}, which we describe in more detail at the end of this section.

\subsection{A Consistency Check for the Tidal Stripping Formation Scenario}
\label{subsec:tidal_stripping}
We can test the plausibility of the stripped nuclei hypothesis by comparing the stellar mass of the clusters to a predicted NSC mass inferred from scaling relationships between black holes, NSCs, and their host galaxies similar to what was done in \citet{Graham2020}. If the luminous GCs with elevated M/L$_{V}$ are the remnant stripped nuclei of galaxies, we expect their stellar mass to be comparable with the NSC mass of their progenitor galaxy \citep{Pfeffer2013}. We therefore use our inferred black hole mass in each cluster to predict the progenitor NSC mass and compare this to the observed cluster stellar mass.  Because of the large scatter in the scaling relations, this test provides only an order of magnitude level consistency check.

We estimate the cluster stellar mass based on their $V$-band luminosity with an assumed M/L$_{V}$ of 1.51 -- the peak lower component of the Gaussian mixture model of Figure~\ref{fig:Hist_ML} -- we call this quantity Mass$_{Pop}$.  We use Mass$_{Pop}$ rather than a virial mass, because if these clusters do indeed have high mass fraction black holes our virial masses over estimate the stellar mass of the clusters.
We then compare this mass to the NSC mass predicted based on the BH mass.  To predict the NSC mass, we first infer the progenitor galaxy stellar mass based on the BH mass -- total galaxy stellar mass from \citet{Reines2015}. A simple rearrangement of  their equation (4) with constants from their equation (5) gives us the following relation between galaxy stellar mass and BH mass:
\begin{equation}
    \label{eq:BH_galaxy}
    \begin{aligned}
& M_{*} = 10^{\frac{\log(M_{BH}) - 7.45\pm0.08}{1.05\pm 0.11} } \times 10^{11}[M_{\odot}]
    \end{aligned}
\end{equation}
Where $M_{*}$ is the galaxy stellar mass, and $M_{BH}$ the BH mass. \citet{Reines2015} report a 0.55 dex scatter around this relation. The $M_{*}$ is then translated back into an expected NSC mass using equation (1) from \citet{Neumayer2020}:

\begin{equation}
    \label{eq:NSC}
    \begin{aligned}
& \log(M_{NSC}) = 0.48\:\log(\frac{M_{*}}{10^{9}M_{\odot}}) + 6.51
    \end{aligned}
\end{equation}

With $M_{NSC}$ being the NSC mass. \citet{Neumayer2020} report a 0.6 dex scatter around this relation. The comparison of the Mass$_{Pop}$ values and the predicted NSC mass are shown in the right panel of Figure~\ref{fig:BHmass}. We calculated the errors in the predicted NSC mass by combining (in quadrature) the scatter of equation \ref{eq:BH_galaxy} and \ref{eq:NSC}.
We find a good agreement between our Mass$_{Pop}$ and predicted NSC masses, with a mean difference of $0.11$~dex for our significant BH predictions (those with 3$\sigma$ M/L$_V$ enhancements; these are shown as solid blue squares).  This agreement is surprisingly good -- the difference is much smaller than the scatter of relations from \citet{Reines2015} and \citet{Neumayer2020}.  We also show data from similar measurements by \citet{Mieske2013} and literature UCDs with dynamical BH masses \citep{Seth2014m60,Ahn2017,Ahn2018,Afanasiev2018} and a recent BH mass estimate from the most massive M31 globular cluster by \citet{Pechetti2021}.  For objects in \citet{Mieske2013} we estimated the NSC mass by multiplying their $V$-band luminosities with their M/L$_{V}$ based on stellar population models for each individual object in \citet{Mieske2013}.  For the dynamical measurements, the stellar mass estimates on the y-axis are dynamically estimated based on the inner components of the UCDs. These show a rough agreement with the predictions with a few outliers including M59-UCD3 ($\log$(Mass$_{Pop})\simeq8.2$) \citep{Ahn2018}.

We also made a similar comparison using the NSC mass -- black hole mass relation derived by \citet{Graham2020}, who previously showed the good agreement between the BH masses and stellar masses for the UCDs with dynamical BH mass estimates. The \citet{Graham2020} relation (their Eq.~8) combines the previous NSC--spheroid \citep{Graham2016} and black hole mass--spheroid \citep{Scott2013} relationships to get a direct correlation between the NSC and BH masses.  
The agreement is less good than using the total stellar mass relations above; we find the Mass$_{pop}$ is on average 0.5 dex lower than the predicted NSC masses using the \citet{Graham2020} relation for our significant BH detections. This discrepancy could be due to the extrapolation of the \citet{Scott2013} relation to lower BH masses than the galaxies it was based on, or simply due to the large scatter in this relationship (black hole masses have a range of $\sim$2 dex at a given spheroid mass). 

Overall, these comparisons support the idea that some of our luminous GCs are the remnant NSCs of stripped galaxies.  We stress here that these inflated mass-to-light ratio objects are likely just a subset of all the stripped nuclei around Cen~A, as stripped nuclei could lack BHs or have BHs with mass fractions too small for us to detect.  

As an alternative to central black holes, the inflated mass-to-light ratios could be due to tidal effects during stripping \citep{Forbes14}. However, two points argue against this: (1)  elevated mass-to-light ratios are seen in nearly a third of our luminous GCs, while inflated mass-to-light ratios due to tidal effects should be limited in duration and thus frequency, and (2) we see no trend of the fraction of luminous GCs with inflated mass-to-light ratios with radius.  We find a mean galactocentric radius of $21.2$~kpc and $21.8$~kpc for luminous GCs with inflated and  non-inflated M/$L_V$ respectively.  

Another possibility is that the inflated mass-to-light ratios are due to variations from the canonical initial stellar mass function (IMF) in luminous GCs. Both top-heavy IMFs that would produce large numbers of stellar remnants \citep{Dabringhausen2009} and bottom-heavy IMFs that would produce an abundance of low-mass stars \citep{Mieske2008} could explain the apparent high mass-to-light ratios. However, the dynamical detection of central SMBHs in massive UCDs \citep{Seth2014m60,Ahn2017,Ahn2018,Afanasiev2018} argues against these interpretations, as they show the mass-to-light ratio clearly rises towards the center, and the outer mass-to-light ratio are consistent (or lighter than) predictions from stellar population models with standard IMFs.  
Additionally, a metallicity and density dependent IMF combined with a mass dependent BH retention fraction has been proposed to explain the M/$L_V$ seen in UCDs  \citep{Jerabkova2017,Mahani2021}. The mass segregation of the stellar mass BHs can produce a compact sub-cluster of BHs in lower mass UCDs such as those we discuss here that raises their inferred M/$L_V$, although this mechanism cannot explain the central dispersion rises in the most massive UCDs ($>$10$^8$~M$_\odot$).  Related work by \citet{Kroupa2020} suggests that a merger of the stellar mass BHs to form a central massive SMBH can occur at the centers of galaxies with spheroid masses $\gtrsim 10^9$~M$_\odot$; this threshold is close to the inferred spheroid masses in our sample of luminous GCs.
\subsection{\textit{VHH81-01}, a Very Massive Cen~A Cluster} 
\label{subsec:posa_20}
The cluster \textit{VHH81-01} has the highest mass-to-light ratio in our sample with M/L$_{V}=7.16^{+1.16}_{-1.0}$, and also the highest predicted black hole mass. The virial mass estimate is 1.25$\times$10$^7$~M$_\odot$, nearly identical to the Cen~A cluster HCH99~18 from \citep{Rejkuba2007}, with a M/L$_{V}=4.7^{+1.2}_{-1.6}$ (which they also suggest is a stripped galaxy nucleus). \textit{VHH81-01} has one of the two highest
virial mass estimates of all clusters in Cen~A.

\textit{VHH81-01} has literature velocity dispersion measurements from several sources \citep{Rejkuba2007,Taylor2010,Taylor2015}. The \citet{Rejkuba2007} UVES dispersion is shown in Figure~\ref{fig:Vdisp_lit} and is 4.6~km/s lower than our derived dispersion of 17.6$\pm$0.8 km/s. While the UVES data \citet{Rejkuba2007} is higher resolution, their global dispersion of $12.4\pm0.8$~km/s agrees within the 1~sigma errors with our global dispersion of 14.8$\pm$1.8~km/s. Our derived dispersion agrees well with the values from \citet{Taylor2010} 15.9$\pm$1.6 km/s and \citet{Taylor2015} 17.3$\pm$0.8 km/s.   If we assume the $12.4\pm0.8$~km/s global velocity dispersion reported in \citet{Rejkuba2007} as the true value, we still get a high a mass-to-light of  M/L$_{V}=4.91^{+0.63}_{-0.48}$ and a predicted BH mass of $4.96^{+1.51}_{-1.2}\times10^{5}$~M$_{\odot}$, thus giving qualitatively similar results.

\begin{figure}
\plotone{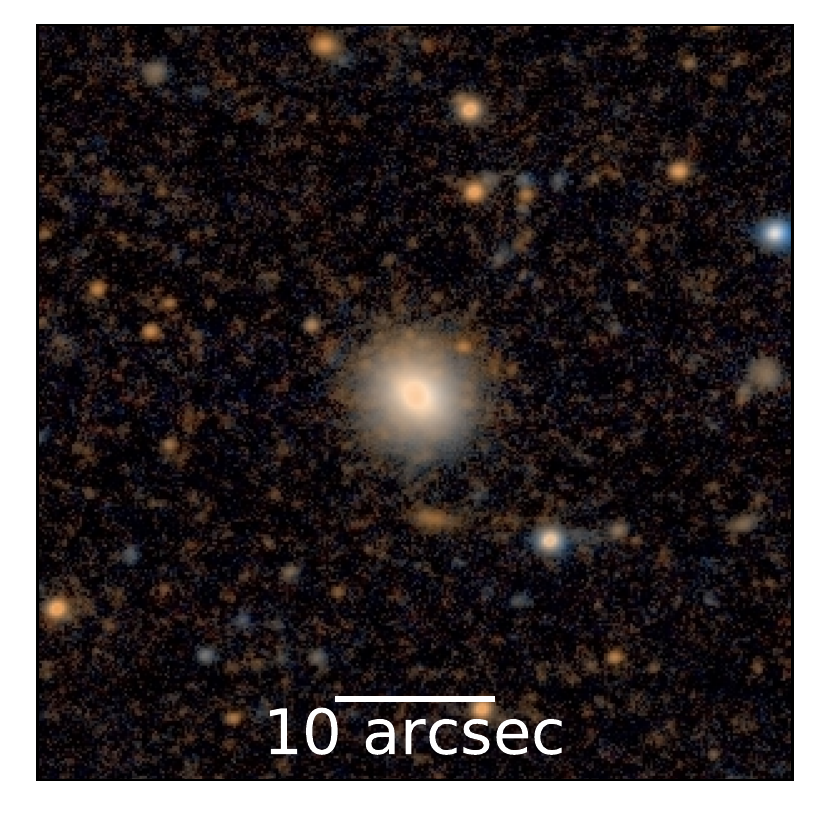}
\caption{Color cutout of \textit{VHH81-01} from ground-based PISCeS imaging at the $6.5$~m telescope Magellan Clay \citep{Crnojevic2016}. The cutout is $0.8\arcmin\times$ $0.8\arcmin$ in size, and oriented such that north is up and east is left. \label{fig:PISCeS_cutout}}
\end{figure}

Our effective radius estimate of $1.7$\arcsec~(24.4~pc) derived from ground-based PISCeS imaging data (see Figure~\ref{fig:PISCeS_cutout}) is in good agreement with that from \citet{Taylor2015}, which is from unpublished ground-based analysis by M. Gom\'ez. Given the good seeing and large effective radius, we do not think that our size estimate could be falsely inflating the mass-to-light ratio. Furthermore we do not observe an unusual $g-r$ color for \textit{VHH81-01} that could underestimate its $L_V$ and produce an inflated M/$L_V$.  Therefore, it appears this object truly has a high mass-to-light ratio and is the most likely of all of our objects to host a high mass fraction black hole.

In addition to \textit{VHH81-01}, there are two other objects in \citet{Rejkuba2007}, \textit{HGHH92-C11} and \textit{HGHH92-C21}, that have similarly high M/L$_{V}$ to \textit{VHH81-01}. All three of these objects are highly flattened, with ellipticity $\gtrsim$0.25 (greater than any MW globular clusters). We know that nuclear star clusters are often quite flattened, both in late-type galaxies and in massive early types \citep{Seth2006,Spengler2017,Neumayer2020}. \textit{HGHH92-C21} has an upper limit in the black hole mass of $1\times 10^{6} M_{\odot}$ from \citet{Karina2018} (published there as UCD320), corresponding to BH mass fraction of $38\%$. \citet{Karina2018} also found a much lower M/L than previously published work due to a smaller derived effective radius for  \textit{HGHH92-C21}.

\section{Conclusions\label{sec:conclusion}}

In this paper we present new high-resolution spectral data from M2FS and MIKE for 321 luminous GC candidates around Cen~A. Of the 321 luminous GC candidates we can reliably determine the radial velocities for 219. Based on their radial velocities, we discover 78 luminous GCs are members of Cen~A, of which 27 are new discoveries.  Of these 78, we can reliably measure the velocity dispersion of 57, including first velocity dispersion measurements of 48 clusters.   
This more than doubles the sample of reliable velocity dispersion measurements, and represents measurements of nearly half of all luminous GCs around Cen~A.


To determine radii for these clusters, we updated the BP-RP {\em Gaia} excess factor and half-light radii relation from \citet{KV2020} using {\em Gaia} EDR3 \citep{GaiaEDR3_phot}.  {\em Gaia} EDR3 is more complete than {\em Gaia} DR2 used in \citet{KV2020} and the BP-RP excess factor has been recalculated.  This revised relation (eq.~\ref{eq:1}) can be used to estimate half-mass radii for extragalactic globular clusters with sizes $< 1\arcsec$ to an accuracy of $14$\%.

We combine size measurements  with the derived  velocity dispersions to determine virial masses and $V$-band mass-to-light ratio for 57 luminous GCs. We find a bimodal distribution of M/L$_V$ with a second population of elevated mass-to-light ratios (M/L$_V>2.3$).  Using a two component Gaussian mixture model we find that $35$\% (20/57) of our luminous GC have inflated M/L$_V$. 

Our preferred explanation for the elevated M/L$_V$ in these clusters is the presence of high mass fraction central black holes in the clusters
. We create JAM dynamical models based on the integrated velocity dispersion to derive the central black hole mass needed to produce the elevated mass-to-light ratio. We find that black hole masses comprising $4-18$\% of the luminous GC mass can explain the elevated mass-to-light ratios in these clusters. The maximum inferred BH mass is $8.61^{+10.3}_{-0.12}\times10^{5}$~M$_{\odot}$ for the cluster \textit{VHH81-01}. 
The possible presence of high mass fraction black holes supports the idea that some or even all of the 20 luminous
GCs with elevated M/L$_V$ are the remnant nuclear star clusters of stripped galaxies.  Future high spatial resolution kinematic observations will be able to verify directly if these clusters contain high mass fraction black holes.


\section*{Acknowledgement}

Work by A.D.~and A.C.S.~has been supported by NSF AST-1813609.  AKH and DJS acknowledge support from NSF grants
AST-1821967 and 1813708.  NC acknowledge support by NSF grant AST-1812461. JS acknowledges support from NSF grant AST-1812856 and the Packard Foundation. We thank Christian I.~Johnson for his recipe book for reducing M2FS data. We also thank Paul Martini for providing the MIKE spectroscopic data for 14 luminous GCs from his 2004 paper, and Pauline Barmby for globular cluster structural properties from \citet{McLaughlin2008}.

This paper made use of the the MGE fitting method and software by \citet{MultiGaussian_Cappellari} and JAM modelling method of \citet{JAM_cappellari}, as well as the \rm astropy package \citep{Astropy2013}.

%

\vspace{5mm}
\facilities{Magellan:Clay (LDSS2 imaging spectrograph)}




\bibliography{References}
\bibliographystyle{aasjournal}

\appendix

\section{HST V-band Magnitudes}
\label{app:magnitudes}

We found five objects (\textit{T17-1498},~\textit{T17-1511},~\textit{KV-280},~\textit{KV19-289},~\textit{T17-1614}) where the magnitudes had large differences between their (very bright) NOAO DR2 values and (fainter) {\em Gaia} magnitudes (i.e.~$|$g-G$| > 2$). PISCeS imaging cutouts for these objects show they are in crowded or star-forming regions that affect the ability to obtain reliable aperture photometry as is used for the NOAO catalog.
Fortunately, all five objects had available HST $F555W$ imaging -- we obtained $V$-band magnitudes from the HST Legacy Archive SExtractor catalogs $F555W$ SExtractor imaging photometry. As these magnitudes are Vega based, the $F555W$ magnitude we assume this magnitude is equal to the $V$ band magnitude (with typical $V-F555W$ from \citet{PARSEC-COLIBRI} models of 0.02 to 0.04 for old stellar populations). Additionally, three luminous GCs have  $g-r$ colors not consistent with the rest of our luminous GCs. ~\textit{KV19-273} and ~\textit{KV19-295} have very red colors, $g-r>1.1$, and ~\textit{T17-1648} a very blue color of $g-r\sim 0.1 mag$. PISCeS imaging cutouts for these objects show that two are in crowded regions and one is close to a bright star. Only \textit{KV19-273} has available HST $F555W$ imaging and we were able to obtain $V$-band magnitude as described above. For ~\textit{KV19-295} and ~\textit{T17-1648} their bad $g-r$ color impacts our ability to estimate $V$-band magnitudes, and their $V$-band magnitudes may be wrong.
Two more luminous GCs, \textit{Fluffy} and \textit{vhh81-5} and a nuclear star cluster \textit{ESO269-06} were not present in {\em Gaia} EDR3 source catalog and we estimated their sizes from fits to HST imaging data (see Sec.~\ref{subsec:radii}). To be consistent, we estimated their $V$-band magnitude from the same data assuming V - $F606W$ $\sim 0.25$ based on expected colors for metal-poor old stellar population using Padova SSP models \citep{PARSEC-COLIBRI}. Finally, we use the $V$-band magnitude of the nuclear star cluster of \textit{KK197-NSC} from \citet{Georgiev2009}. We correct for foreground extinction for these ten sources using the same recipe as that for the NOAO DR2 photometry (see \S\ref{subsec:sample_selection}).

\setcounter{figure}{0}  
\renewcommand\thefigure{A\arabic{figure}} 

\setcounter{table}{0}
\renewcommand{\thetable}{A\arabic{table}}

\startlongtable
\begin{deluxetable*}{llrrrrrrrrrrr}
\tabletypesize{\scriptsize}
\tablecaption{ List of all observed luminous GCs candidates \label{tab:all_GC_data}} 
\tablehead{ \colhead{ID} & \colhead{ID Ref} &\colhead{RA} & \colhead{DEC} & \colhead{S/N} &    \colhead{$g_{0}$} &  \colhead{R.V} & \colhead{R.V err} & \colhead{Metalicity} & \colhead{Metalicity err}  & \colhead{setup} & \colhead{Cen~A?} & \colhead{Ref}\\
 \colhead{} & \colhead{} & \colhead{[deg]} & \colhead{[deg]} & \colhead{} &    \colhead{[mag]} &  \colhead{[km/s]} & \colhead{[km/s]} &  \colhead{[Fe/H]} &  \colhead{[Fe/H]} &  \colhead{} & \colhead{} }
\startdata
  GC\_cand\_37 &        This Work & 200.926244 & -43.565931 &      8.08 & 19.56 & \nodata & \nodata & \nodata &      \nodata &  M2FS ca &  \nodata&  \nodata \\
  GC\_cand\_38 &        This Work & 200.959920 & -43.543763 &      4.68 & 21.04 & \nodata & \nodata & \nodata &      \nodata &  M2FS ca &  \nodata&  \nodata \\
    T17-1027 &       Taylor2017 & 200.969649 & -43.318745 &      4.91 & 20.01 &      502.0 &        7.1 &        0.20 &             0.34 &  M2FS ca &    Y &      New \\
   AAT301956 &      Beasley2008 & 200.977085 & -43.333630 &      5.54 & 20.37 & \nodata & \nodata & \nodata &      \nodata &  M2FS ca &  \nodata&  \nodata \\
  H21-200841 &        This Work & 200.979670 & -43.463985 &      7.23 & 20.51 & \nodata & \nodata & \nodata &      \nodata &  M2FS ca &  \nodata&  \nodata \\
   GC\_cand\_0 &        This Work & 200.989659 & -42.868971 &     18.46 & 18.79 &        0.7 &        5.9 &        1.03 &             0.32 &  M2FS ca &    N &      New \\
   AAT101931 &      Beasley2008 & 200.993959 & -42.954726 &     13.16 & 19.84 &      337.1 &        6.1 &        1.31 &             0.36 &  M2FS ca &    Y &       RV \\
   AAT102120 &      Beasley2008 & 200.998330 & -42.922039 &     14.77 & 19.88 &      254.4 &        5.9 &        1.49 &             0.33 &  M2FS ca &    N &       RV \\
\enddata
\tablecomments{Column 1 shows the Discovery ID for the luminous GC candidates. For 133 targets that were based on preliminary versions of \citet{KV2020,Hughes2021} and do not have literature names, we name them as \textit{GC\_cand} with numbers running from 1 to 133. Literature reference for the target ID from \citet{Bergh1981,Lauberts1982,Hesser1986,Karachentseva1998,Harris2002,Peng2004,Woodley2007,Georgiev2009,Harris2012,Denja2014,Denja2016,Taylor2017,KV2020,Hughes2021}.Column 3 \& 4 shows Right Ascension and Declination in J2000 respectively. The signal-to-noise is shown in Column 5. Column 6 shows the extinction corrected NOAO DR2 g-band magnitude for the target . Column 7 \& 8 shows the heliocentric radial velocity and its error respectively. Column 9 and 10 shows the derived luminosity-weighted metallicity and its error. Column {11} denotes the instrument and setup used in the observation, with ``ca'' \& ``mg'' being Calcium \& Magnesium triplet respectively. For targets observed with MIKE, only the Calcium Triplet measurements are shown in this table. ``Y'' in Column 12 denotes if the target has radial velocity expected for Cen~A globular clusters (RV$>250$~km/s), ``N'' if it has radial velocity consistent with Milky Way foreground stars, and ``N/A'' if we cannot reliably determine its radial velocity. Column 13 shows the literature source of the target. A-E denotes the likelihood rank of being a Cen~A GC from \citet{KV2020}, with A \& E being more and less likely respectively. ``RV'' and ``Sigma'' denotes if the target has previous literature measured radial velocity and velocity dispersion respectively from \citet{Harris2002,Peng2004,martini2004,Rejkuba2007,Taylor2010,Taylor2015}. Finally ``New'' denotes targets that are not in KV20 catalog, and have no previous literature kinematic measurements. }
\end{deluxetable*}

\startlongtable
\begin{deluxetable*}{lrrrrrrrrrrrrr}

\tablecaption{ List of 57 luminous GCs with visually verified velocity dispersion \label{tab:good_GC_data}}
\tablehead{ \colhead{ID} & \colhead{R.V} & \colhead{R.V err} & \colhead{$\sigma$} & \colhead{$\sigma$ err}  & \colhead{$\sigma_{\infty}$}& \colhead{setup} & \colhead{$r_{\rm hm}$} & \colhead{$r_{\rm hm}$ Ref}& \colhead{M$_{V}$} & \colhead{E(B-V)} & \colhead{M/L$_{V}$}& \colhead{BH Mass}&\colhead{Ref}\\
 \colhead{} &  \colhead{[km/s]} & \colhead{[km/s]} & \colhead{[km/s]}& \colhead{[km/s]}&\colhead{[km/s]}& \colhead{} & \colhead{[\arcsec]} & \colhead{} & \colhead{[mag]} & \colhead{[mag]}& \colhead{}& \colhead{[M$_{\odot}$]}&\colhead{}}
\startdata
      HH-22 & 615.2 &      5.93 & 13.07 &        1.82 &      12.44 &  M2FS ca & 0.2498 &       Gaia & -10.18 &       0.108 & 1.25$^{0.44}_{0.36}$ & \nodata &       RV \\
   T17-1253 & 548.3 &      5.94 & 11.95 &        1.49 &      11.66 &  M2FS ca & 0.1549 &       Gaia &  -8.66 &       0.106 & 2.77$^{+0.89}_{-0.73}$ &  1.12$^{+0.9}_{-0.6}\times10^{5}$ &      New \\
   HHH86-14 & 699.9 &      5.92 & 15.48 &        1.91 &      14.12 &  M2FS ca & 0.3407 &    M08/ACS & -10.39 &       0.109 & 1.81$^{+0.79}_{-0.49}$ & \nodata &       RV \\
   HHH86-30 & 780.2 &      5.91 & 21.97 &        4.69 &      20.97 &  M2FS ca & 0.2352 &       Gaia & -11.02 &       0.106 & 1.54$^{+0.82}_{-0.61}$ & \nodata &       RV \\
      K-034 & 460.6 &      5.91 & 14.00 &        2.19 &      13.15 &  M2FS ca & 0.3001 &       Gaia & -10.53 &       0.102 & 1.21$^{+0.47}_{-0.39}$ & \nodata &       RV \\
  PFF-gc028 & 539.6 &      5.92 & 10.20 &        1.48 &       9.90 &  M2FS ca & 0.1768 &       Gaia &  -9.40 &       0.101 & 1.15$^{+0.41}_{-0.34}$ & \nodata &       RV \\
   HHH86-28 & 468.1 &      5.97 & 13.35 &        1.27 &      12.68 &  M2FS ca & 0.2557 &       Gaia &  -9.75 &       0.122 & 1.98$^{+0.54}_{-0.43}$ & \nodata &       RV \\
  HGHH-G066 & 581.4 &      5.93 & 13.63 &        1.68 &      13.25 &  M2FS ca & 0.1692 &       Gaia &  -9.49 &       0.101 & 1.82$^{+0.57}_{-0.47}$ & \nodata &       RV \\
\enddata
\tablecomments{Column 1-3 are identical to columns 1,6 and 7 in Table~\ref{tab:all_GC_data}.Columns 4, 5 and 6 show the derived  velocity dispersion, its error and the global velocity dispersions (See \S\ref{sec:M/L}) respectively. In Column 7 are the calculated the half-mass radii in arcseconds. Column 8 denotes the source of the calculated half-mass radii, with ``{\em Gaia}'' being the BR$_{excess}$-half-mass radii relation described in Section~\ref{sec:radii_lumin}, and M08/ACS and M08/STIS are sizes from \citet{McLaughlin2008}. Column 9 shows the extinction corrected absolute $V$-band magnitude (assuming a distance modulus of $(m-M)_{V}=27.91$). Column 10 shows the Galactic dust reddening used in column 9. Column 11 shows the calculated $V$-band mass-to-light ratio. Column 11 shows the predicted black hole mass necessary for luminous GCs with M/L$_{V}\geq2.3$ to reproduce the observed elevated M/L$_{V}$. Column 13 is identical to column 13 in Table~\ref{tab:all_GC_data}. We highlight ($^{*}$) two luminous GCs with probably incorrect $V$-band luminosity (See Section~\ref{sec:radii_lumin}) }
\end{deluxetable*}

\begin{deluxetable*}{l|rrrrrr}
\tablecaption{List of updated mass-to-light ratios for literature globular clusters\label{tab:Rejkuba}}
\tablewidth{0.5\columnwidth} 
\tablehead{ \colhead{ID} & \colhead{RA} & \colhead{DEC} & \colhead{$\sigma$} & \colhead{Mv} & \colhead{$r_{hm}$} & \colhead{M/Lv} \\
\colhead{} & \colhead{[deg]} & \colhead{[deg]} & \colhead{[km/s]} & \colhead{[mag]} & \colhead{[pc]} }
\startdata
  HGHH92-C7 &  201.522542 & -42.942333 & 21.10 & -11.09 &  6.82 & 2.22$^{+0.07}_{-0.07}$  \\
 HGHH92-C11 &  201.228042 & -43.022694 & 19.20 & -10.35 & 10.67 & 5.68$^{+0.80}_{-0.75}$  \\
 HGHH92-C12 &  201.273833 & -43.175194 & 13.10 & -10.52 &  7.05 & 1.49$^{+0.19}_{-0.17}$  \\
 HGHH92-C17 &  201.415542 & -42.933111 & 20.90 & -10.63 &  7.41 & 3.61$^{+0.64}_{-0.58}$  \\
 HGHH92-C21 &  201.469750 & -43.096222 & 19.00 & -10.39 &  9.19 & 4.62$^{+0.48}_{-0.47}$  \\
 HGHH92-C22 &  201.473208 & -42.985444 & 17.90 & -10.11 &  4.99 & 2.88$^{+0.18}_{-0.17}$  \\
 HGHH92-C23 &  201.477417 & -42.990389 & 30.50 & -11.66 &  4.46 & 1.79$^{+0.14}_{-0.13}$  \\
 HGHH92-C29 &  201.168292 & -43.301472 & 16.10 & -10.11 &  6.68 & 3.12$^{+0.47}_{-0.42}$  \\
\enddata
\tablecomments{Column 1 shows the name of the GC as identify in \citet{Rejkuba2007}. Column 4 shows the derived  velocity dispersion from \citet{Rejkuba2007}. Column 5 shows the extinction corrected $V$-band absolute magnitude from \citet{Rejkuba2007}. Column 6 shows the corresponding half-mass radii using transformation \ref{eq:rh_to_rhm} with half-light radii from \citep{McLaughlin2008}. Column 7 shows the updated $V$-band mass-to-light ratio. We calculated the mass-to-light ratio by taking the ratio of virial mass from equation~\ref{eq:2} using a central velocity dispersion based in values of Column 4 with the $V$-band luminosity from Column 5. For the central velocity dispersion we assumed a 6\% correction between values of column 4 and $\sigma_{\infty}$ based on the average correction by \citet{Rejkuba2007}).}
\end{deluxetable*}

\begin{figure*}[h!]
\plottwo{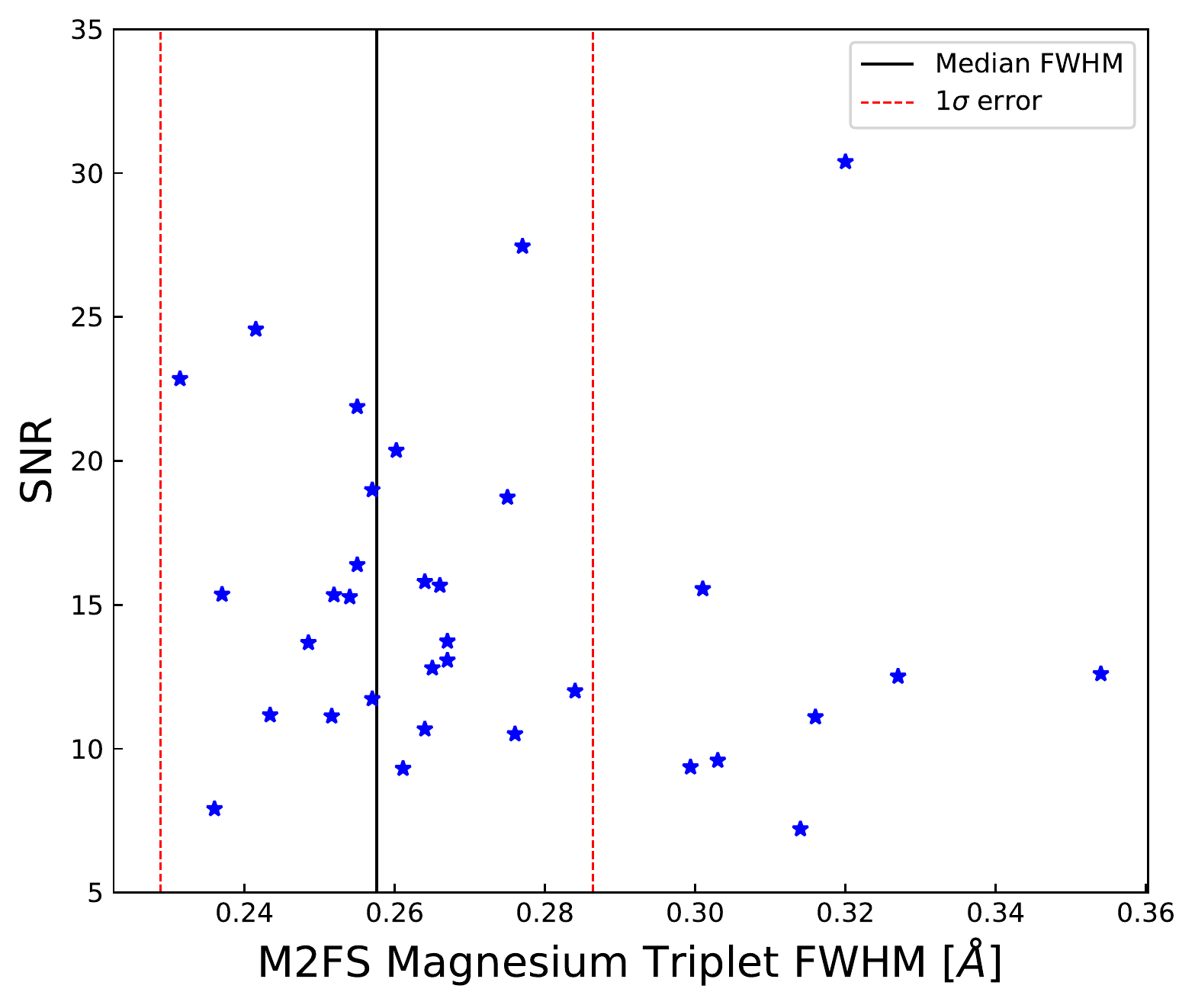}{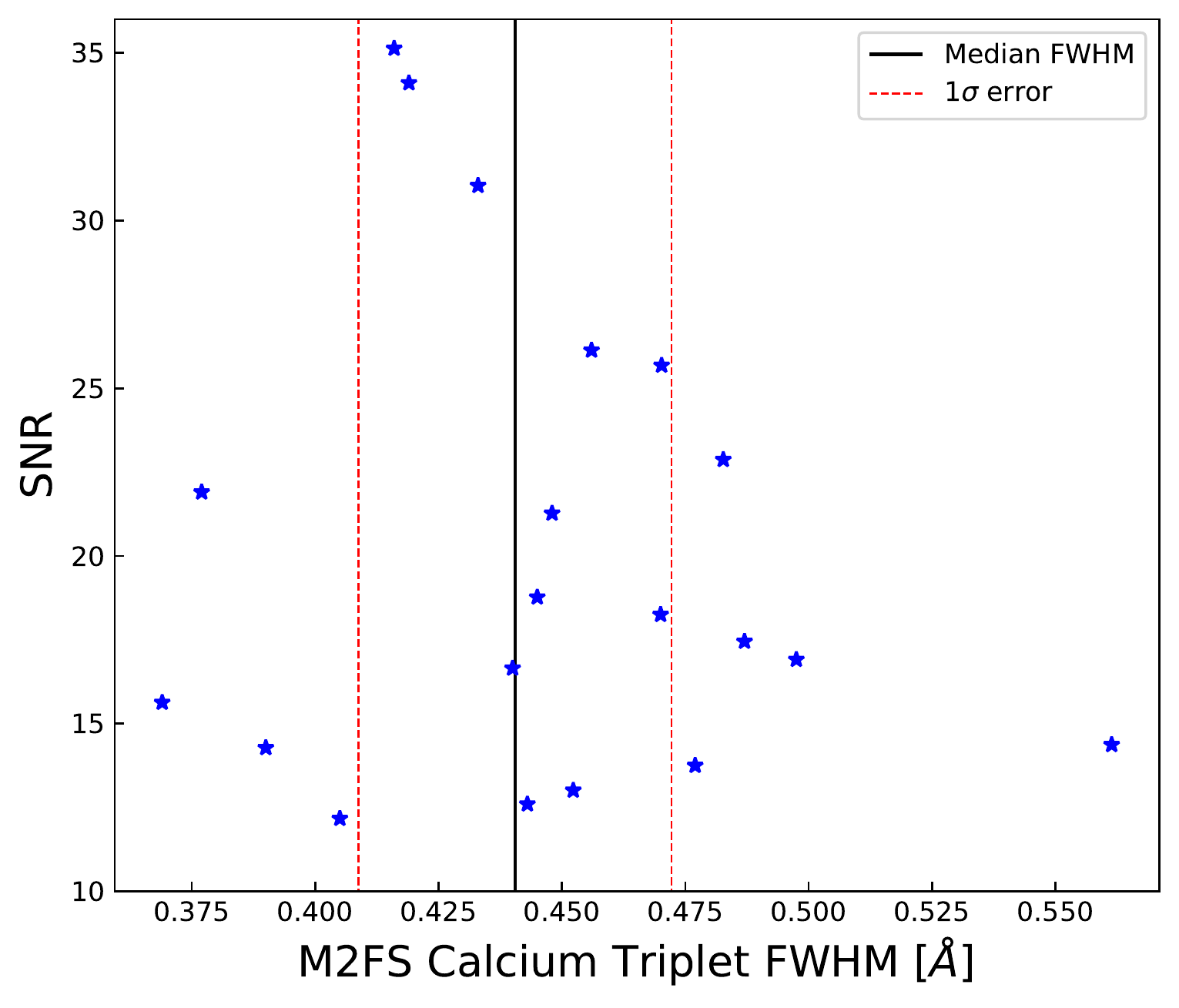}
\caption{ Line-Spread Function determination for Mg \& Ca M2FS data. We restricted our Line-Spread Function determination for the stars in our field with S/N $\geq\:20$.    \label{fig:LSF_lot}}
\end{figure*}

\end{document}